\documentclass[iop]{emulateapj}
\usepackage{amsmath}
\usepackage{rotating}
\usepackage[colorlinks=true,citecolor=blue]{hyperref}

\slugcomment{\today}  %Submitted to ApJ on
\begin{document}
%\begin{CJK*}{UTF8}{gbsn}
\title{The physical properties of $Fermi$-4LAC flat spectrum radio quasars}
\author{Can Tan\altaffilmark{1}, Rui Xue\altaffilmark{2, 3, 4}, Lei-Ming Du\altaffilmark{1}, Shao-Qiang Xi\altaffilmark{2, 3}, Ze-Rui Wang\altaffilmark{2, 3, 5}, and Zhao-Hua Xie\altaffilmark{1}}
\altaffiltext{1}{Department of Physics, Yunnan Normal University, 650500, Kunming, China}
\altaffiltext{2}{School of Astronomy and Space Science, Nanjing University,210093,Nanjing, China}
\altaffiltext{3}{Key laboratory of Modern Astronomy and Astrophysics (Nanjing University), Ministry of Education, Nanjing 210023, People's Republic of China}
\altaffiltext{4}{Corresponding author, email: dg1726013@smail.nju.edu.cn}
\altaffiltext{5}{Corresponding author, email: zerui\_wang62@163.com}

\begin{abstract}
In this work, we collect quasi-simultaneous infrared, optical, X-ray and $\gamma$-ray data of 60 $Fermi$-4LAC flat spectrum radio quasars (FSRQs). In the framework of the conventional one-zone leptonic model, we investigate the physical properties of $Fermi$-4LAC FSRQs' jets by modeling their quasi-simultaneous spectral energy distributions (SEDs). Our main results are summarized as follows. (1) There is a linear correlation between synchrotron peak frequency and curvature of the electron energy distribution. As suggested by previous works, the slope of the best linear fitting equation of this correlation is consistent with statistic acceleration which needs a fluctuation of fractional acceleration gain. (2) The gamma-ray dissipation regions are located at the range from 0.1 to 10 pc away from the super-massive black hole, and located outside the broad-line region (BLR) and within the dusty torus (DT). (3) A size relation $P_{\rm e}$ (the kinetic power carried in relativistic electrons) $\sim$ $P_{\rm B}$ (Poynting flux) $\leq$ $P_{\rm r}$ (the radiative power ) $<$ $P_{\rm p}$ (the kinetic power in cold protons) is found in our modeling. Among them, $P_{\rm e}\sim P_{\rm B}$ suggests that SEDs of almost all FSRQs with parameters are close to equipartition between the magnetic field and the relativistic electrons. The $P_{\rm e} < P_{\rm r}$ suggest that the most energy of the relativistic electrons are dissipated by EC radiation for FSRQs. (4) There is an anti-correlation between the peak energy of SEDs ($\gamma_{\rm peak}$) and the jet power ($P_{\rm jet}$), which is consistent with the blazar sequence.
\end{abstract}


\section{Introduction}
Blazars are the most extreme and powerful subclass of active galactic nuclei (AGNs), and their observed emission is dominated by a relativistic jet pointed in the direction of the observer \citep[][]{1995PASP..107..803U}. They have high luminosity, large amplitude and rapid variability, high and variable polarization, radio core dominance, and apparent super-luminal speeds \citep{1995PASP..107..803U,2015A&ARv..24....2M}. Blazars are divided into BL Lacertae objects (BL Lacs) with weak or no emission lines  ($\rm EW$ $<$ $\rm 5\AA$) and flat spectrum radio quasars (FSRQs) with stronger emission lines $\rm EW \ge 5\AA$ \citep{1995PASP..107..803U}. The spectral energy distribution (SED) of blazars are characterized by two bumps in the $\rm log\nu-log\nu F_{\nu}$ diagram. The low energy bump at the IR-optical-UV band is well explained by synchrotron emission from relativistic electrons, and the high energy bump at the GeV-TeV gamma-ray band is attributed to the inverse Compton (IC) scattering \citep[e.g.,][]{1995ApJ...446L..63D, 2002ApJ...575..667D, 2007Ap&SS.309...95B}. The seed photons for IC process could be from the synchrotron radiation (synchrotron self-Compton, SSC; \citep[e.g.,][]{1992ApJ...397L...5M,1998ApJ...509..608T} or from external photon fields (external-Compton, EC), such as the accretion disk  \citep[e.g.,][]{1993ApJ...416..458D}, the broad-line region  \citep[BLR; e.g.,][]{1994ApJ...421..153S,2006ApJ...646....8F}, and the dusty torus  \citep[DT; e.g.,][]{2000ApJ...545..107B,2002A&A...386..415A,2005ApJ...629...52S}. Also, the hadronic model is an alternative explanation for the high-energy emission \citep[e.g.,][]{1993A&A...269...67M,2003APh....18..593M,2012ApJ...755..147D,2012A&A...546A.120D, 2019ApJ...886...23X}. The modeling of SEDs allows us to investigate the intrinsic physical properties of emitting region in the jet \citep[e.g.,][]{2008MNRAS.387.1669G,2009MNRAS.397..985G,2010MNRAS.409L..79G,2014ApJ...788..104Z,2014MNRAS.439.2933Y,2017MNRAS.464..599D, 2018ApJS..235...39C}.

\citet{1986ApJ...308...78L} find an anti-correlation between synchrotron peak frequency and its curvature. This correlation can be explained in terms of the acceleration processes of electrons \citep{2004A&A...413..489M,2011ApJ...739...66T}. \citet{2011MNRAS.417.1881R} and \citet{2017MNRAS.464..599D} check this correlation by modeling the SEDs. Based on previous studies \citep{2004A&A...413..489M,2011ApJ...739...66T}, \citet{2014ApJ...788..179C} deduce the theoretical relationship between synchrotron peak frequency and its curvature for different acceleration mechanisms. For stochastic acceleration, statistical acceleration with energy-dependent acceleration, and fractional acceleration gain, the predicted theoretical value of slope $k$ ($1/b_{\rm syn} = k {\rm log}\nu_{\rm p} + m$, where $b_{\rm syn}$ is the synchrotron curvature, ${\rm log}\nu_{p}$ is the logarithm of the synchrotron peak frequency) is 2, 5/2, and 10/3, respectively. The result of \citet{2014ApJ...788..179C} is $k$ = 2.04 $\pm$ 0.03, which favors stochastic acceleration mechanism. \citet{2016MNRAS.463.3038X} collected a large sample of BL Lacs and FSRQs. The slope of BL Lacs ($k$ = 1.87 $\pm$ 0.19) is consistent with the result of \citet{2014ApJ...788..179C}. On the other hand, the slope of FSRQs ($k$ = 3.69 $\pm$ 0.24) is closely consistent with the statistic acceleration for the case of fluctuation of fractional acceleration gain \citet{2014ApJ...788..179C}.

The location of the dissipation regions, where the bulk energy of the jet is converted to a distribution of highly relativistic particles, can be a clue to understand the origin of ambient photon fields in the jets \citep[e.g.,][]{2011ApJ...726L..13A,2012ApJ...758L..15D,2014ApJ...789..161N,2016ApJ...821..102B}. However, there is no consensus on the location of the dissipation region. \citet{2010ApJ...717L.118P} shows that the sharpness and position of $\gamma$-ray breaks can be reproduced by the absorption via photon-photon pair production on He{\footnotesize II} Lyman recombination continuum and lines. Their result implies that the location of the emission region is close to the super-massive black hole (SMBH). On the other hand, the new $\gamma$-ray data from $Fermi$-LAT and the currently operational Cerenkov telescopes indicate that the dissipation region is actually quite remote from SMBH \citep[e.g.,][]{2016ARA&A..54..725M}. There are two main diagnostics for the energy dissipation region: (i) the variability timescales \citep[e.g.,][]{2010ApJ...712..957A,2010ApJ...715..362J,2010MNRAS.405L..94T,2011MNRAS.418...90L,2012JPhCS.355a2032A,2012ApJ...751L...3G,2013MNRAS.431..824B,2015MNRAS.452.1280R}, and (ii) fitting the SEDs \citep[e.g.,][]{2009ApJ...692...32D,2009MNRAS.397..985G,2012arXiv1202.6193G,2012MNRAS.421.2956Z,2014ApJS..215....5K,2017ApJS..228....1Z,2018ApJ...859..168Y}. Since the variability timescale only implies the size of dissipation region but not at any particular location, we consider the SED argument to study the location of dissipation region in this work \citep[e.g.,][]{2014ApJ...790...45P,2015MNRAS.454.1310Y,2018ApJ...852...45W}.

The $Fermi$-4LAC source catalogue was released recently \citep{2019arXiv190210045T}. With the new gamma-ray data, we fit the SEDs of 60 $Fermi$-4LAC FSRQs which have contemporaneous multi-wavelength observation data to analyze the properties of the jets, trace the location of the dissipation regions, and explore the particle acceleration mechanisms in the framework of the conventional leptonic EC model. This paper is organized as follows: In Sect.~\ref{sample}, we present the sample, the model description is presented in Sect.~\ref{model}. Then, results and discussions are shown in Sect.~\ref{RD}. Finally, we end with a conclusion of this work in Sect.~\ref{con}. The cosmological parameters $H_{0}=70\ \rm km\ s^{-1}Mpc^{-1}$, $\Omega_{0}=0.3$, and $\Omega_{\Lambda}$= 0.7 are adopted in this work.

\section{The Sample}\label{sample}
We collect a sample with 60 FSRQs from the forth LAT AGN catalogue (4LAC). Their broadband SEDs from infrared to $\gamma$-ray band are available based on (quasi-) simultaneous observations. Since the radio data cannot be explained by one-zone leptonic model and most of them on the SED Builder are collected around the 1990s, we do not consider the radio data. First, we obtain all data and corresponding observation times in the SSDC SED Builder, an online service developed at the Space Science Data Center \citep{2011arXiv1103.0749S}\footnote{http://tools.ssdc.asi.it/SED/}. Second, we search for the intersection of observation time from infrared to X-ray band, and ensure that the interval of observation time between any two bands shall not exceed 7 days. We obtain the $Fermi$ GeV data by integrating 2 months\footnote{see Appendix A for more details} that include the previous time intersection, because the gamma-ray data on the SSDC website are all collected by annual integration and this website has not yet collected the data of $Fermi$-4LAC. Finally, we collect the quasi-simultaneous data from infrared to $\gamma$-ray band. The data from infrared to X-ray band are observed within one week, and the $Fermi$-LAT data are integrated two months. Therefore, the multi-frequency data in our sample are quasi-simultaneous, but not really simultaneous.

Details are presented in Table~\ref{table1}, where column(1) gives the name of the Fermi catalogue, column(2) gives the source name, column(3) gives the right ascension, column(4) gives the declination, column (5) gives the observation time between infrared and X-ray band, and column(6) gives the integration time of $\gamma$-ray.

For the broad-line luminosity of 44 FSRQs in our sample, we select these values from \citet{2009MNRAS.397..985G} and \citet{2016MNRAS.463.3038X}. For FSRQs without a measured broad-line luminosity, we use  the mean value of $10^{44.87}$ $\rm erg\ s^{-1}$ in our sample.

\section{The Model Description}\label{model}
In this section, all quantities are measured in the comoving frame, unless specified otherwise. We adopt the conventional one-zone EC model to fit the SEDs, which is widely used to study the properties of blazars \citep[e.g.,][]{2010MNRAS.402..497G,2017MNRAS.470.3283S,2018A&A...616A..63A}. The dissipation region is assumed to originate from a single spherical region with radius $R$, which is composed of a plasma of relativistic electrons and cold protons in a uniformly entangled magnetic field $B$. Because of the beaming effect, the observed radiation is strongly boosted by a relativistic Doppler factor $\delta$\footnote{For the relativistic jet close to the line of sight in blazars with a viewing angle of $\theta \lesssim 1/\Gamma$, we have $\delta=[\Gamma(1-\beta \rm cos\theta)]^{-1}\approx \Gamma$.}. In the framework of leptonic model, since the log-parabolic electron spectrum is helpful for the study of acceleration mechanisms, we assume that the steady-state electron energy distribution is a log-parabolic spectrum \citep[e.g.,][]{1962SvA.....6..317K,2004A&A...413..489M,2006A&A...448..861M,2009A&A...504..821P,2011ApJ...739...66T, 2019ApJ...878..140L},

\begin{equation}
     N(\gamma)=N({\frac{\gamma}{\gamma_{\rm 0}}})^{-s-r\log({\frac{\gamma}{\gamma_{\rm 0}})}}, \ \gamma_{\rm min}\leq \gamma \leq \gamma_{\rm max},
     \label{eq:quadratic1}
\end{equation}
where $N$ is the normalization constant in units of $1/\rm cm^3$, $\gamma$ is the the electron Lorentz factor, $\gamma_0$ is the reference energy. $s$ is the spectral index, $r$ is the spectral curvature, $\gamma_{\rm min}$ and $\gamma_{\rm max}$ are the minimum and maximum electron Lorentz factors.

After assuming a steady-state electron distribution $N(\gamma)$, we can calculate the synchrotron, SSC and EC emissions from jets. The synchrotron emission coefficients is calculated with
\begin{equation}
    j_{\rm syn}(\nu) = \frac{1}{4\pi} \int N(\gamma)P(\nu, \gamma) d\gamma,
    \label{eq:quadratic2}
\end{equation}
where $\nu$ is the frequency, $P(\nu$, $\gamma$) is the mean emission coefficient for a single electron integrated over the isotropic distribution of pitch angles \citep[e.g.,][]{1988ApJ...334L...5G,2001A&A...367..809K}. And the synchrotron absorption coefficient is calculated with
\begin{equation}
    \alpha_{\rm syn}(\nu) = -\frac{1}{8\pi \nu^2 m_e} \int d\gamma P(\nu, \gamma)\gamma^2 \frac{\partial}{\partial \gamma}[\frac{N(\gamma)}{\gamma ^2}].
    \label{eq:quadratic3}
\end{equation}
 Then we can calculate the synchrotron intensity using the radiative transfer equation
\begin{equation}
    I_{\rm syn}(\nu) = \frac{j_{\rm syn}(\nu)}{\alpha_{\rm syn}(\nu)}[1-\frac{2}{\tau(\nu)^2}(1-\tau e^{-\tau(\nu)}-e^{-\tau(\nu)})],
    \label{eq:quadratic4}
\end{equation}
where $\tau(\nu)= \alpha_{\rm syn}(\nu)2R$ is the optical depth.

The SSC and EC emission coefficients are given as
\begin{equation}
    j_{\rm IC} = \frac{h\epsilon}{4\pi} \int d\epsilon_0 n(\epsilon_0) \int \gamma N(\gamma)C(\epsilon, \gamma, \epsilon_0),
    \label{eq:quadratic5}
\end{equation}
where $\epsilon$ is the scatted photon energy in units of $m_{\rm e}c^2$, ${\epsilon}_{0}$ is the soft photon energy in units of $m_{\rm e}c^2$, $n(\epsilon_0)$ is the number density of seed photons per energy interval and C($\epsilon$, $\gamma$, ${\epsilon}_{0}$) is the Compton kernel given by \citet{1968PhRv..167.1159J}.

The only difference between EC process and SSC process is the origin of seed photons. For SSC process, the seed photons are the synchrotron photons emitted by the same population of electrons. Meanwhile, for EC process, the seed photons are considered to originate from BLR and DT \citep[e.g.,][]{2016MNRAS.463.4469D,2018A&A...616A..63A,2018PhDT........27D}. We can calculate the energy density of BLR ($\mu_{\rm BLR}$) and DT ($\mu_{\rm DT}$) in the jet comoving frame according to \citep{2012ApJ...754..114H}

\begin{equation}
    \mu_{\rm BLR}=\frac{{\eta_{\rm BLR}}{\Gamma}^{2}{L_{\rm d}}}{{3\pi{x_{\rm BLR}^{2}}c[1+(x/x_{\rm BLR})^{3}]}}
    \label{eq:quadratic6}
\end{equation}
and
\begin{equation}
    \mu_{\rm DT}=\frac{{\eta_{\rm DT}}{\Gamma}^{2}{L_{\rm d}}}{{3\pi{x_{\rm DT}^{2}}c[1+(x/x_{\rm DT})^{4}]}},
    \label{eq:quadratic7}
\end{equation}
where $\eta_{\rm BLR}$ = 0.1 and $\eta_{\rm DT}$ = 0.1 are the fractions of the disk luminosity $L_{\rm d}$ reprocessed into BLR and DT radiation, respectively. $x$ is the distance between the position of the dissipation region and the central black hole in the AGN frame. We assume that the characteristic distance of BLR is $x_{\rm BLR}$ = 0.1 ($L_{\rm d}/10^{46}$ ergs$^{-1})^{1/2}$pc and the characteristic distance of DT is $x_{\rm DT}$ = 2.5($L_{\rm d}/10^{46}$ergs$^{-1})^{1/2}$ pc \citep[e.g.,][]{2008MNRAS.387.1669G}. The radiation from both BLR and DT is taken as an isotropic blackbody with a peak at $2\times 10^{15}\Gamma$ Hz \citep{2008MNRAS.386..945T} and $3\times 10^{13}\Gamma$ Hz \citep{2007ApJ...660..117C}, respectively.

Based on the above, we can calculate the IC intensity
\begin{equation}
    I_{\rm IC}(\nu) = j_{\rm IC}(\nu)R,
    \label{eq:quadratic8}
\end{equation}
since the medium is transparent for IC radiation. Finally, the total observed flux density can be calculated by
\begin{equation}
    F_{\rm obs}(\nu_{\rm obs}) = \frac{\pi R^2 \delta^{3}(1+z)}{D_{\rm L} ^2} (I_{\rm syn}(\nu) + I_{\rm IC}(\nu)),
    \label{eq:quadratic9}
\end{equation}
where $D_{\rm L}$ is the luminosity distance, $z$ is the redshift, and ${\nu}_{\rm obs}$ = $\nu \delta/(1+z)$, where $\nu$ is the frequency of photons in comoving frame. Since the very high energy (VHE) $\gamma$-ray photons will be absorbed by the extragalactic background light (EBL), we calculate the absorption in the GeV-TeV band by using the EBL model presented by \citet{2011MNRAS.410.2556D}.

There are 11 free parameters in our model: $R$, $B$, $\delta$, $N$, $\gamma_0$, $\gamma_{\rm min}$, $\gamma_{\rm max}$, $s$, $r$, $x$, and $L_{\rm d}$. Since most of these parameters cannot be directly observed and they are coupled to each other, it will take a long time to reproduce the best SED if we allow all eleven parameters to be free. In this work, we adopt $\gamma_{\rm min} = 48$ \citep{2014ApJ...788..104Z} and $\gamma_{\rm max} = 2\times 10^{6}$ \citep{2005A&A...432..401G}, because our model is not sensitive to these two parameters. In addition, we estimate the disk luminosity through $L_{\rm d} = 10 \times L_{\rm BLR}$ as proposed by \citet{2008MNRAS.387.1669G}.

Based on the above constraints, we can calculate the observed photon spectrum and the corresponding chi-square value $\chi^2$. These parameters are listed in Table~\ref{table2}, and its columns are as follows:
\begin{itemize}
  \item [1.]
  $Fermi$ name;
  \item [2.]
  Source name;
  \item [3.]
  $z$, redshift;
\item [4.]
  $\log L_{\rm BLR}~(\rm erg~s^{-1})$, logarithm of the broad-line luminosity;
\item [5.]
  $R~(10^{17} \rm cm)$, the radius of the dissipation region;
\item [6.]
  $B~(\rm G)$, the magnetic field;
\item [7.]
  $\delta$, the Doppler factors;
\item [8.]
  $s$, the electron spectral index;
\item [9.]
  $r$, the curvature of the electron energy distribution;
\item [10.]
  $N$, the normalization of the electron energy distribution;
\item [11.]
  $\gamma_{0}~(10^2)$, the reference energy;
\item [12.]
  $x~(\rm pc)$, the distance between the dissipation region and SMBH;
\item [13.]
  $\chi^2$, $\chi^2 = \frac{1}{m-dof}\sum_{i=1}^{m}(\frac{\hat {y}_i-y_i}{\sigma_i})^2$, where $m$ is the number of quasi-simultaneous observational data points, $dof$ are the degrees of freedom, $\hat {y}_i$ are the expected values from the model, $y_i$ are the observed data and $\sigma_i$ is the standard deviation for each data point. In our sample, these errors of data points from infrared to X-ray band are collected from the SSDC website. For errors of data points can not be found, we take 1$\%$ of the observed infrared and optical flux and take 2$\%$ of the observed UV and X-ray flux as the errors of these data points \citep[e.g.,][]{2014A&A...567A.135A}.
\end{itemize}

\begin{figure}[htbp]
\centering
\includegraphics[width=1.7\columnwidth]{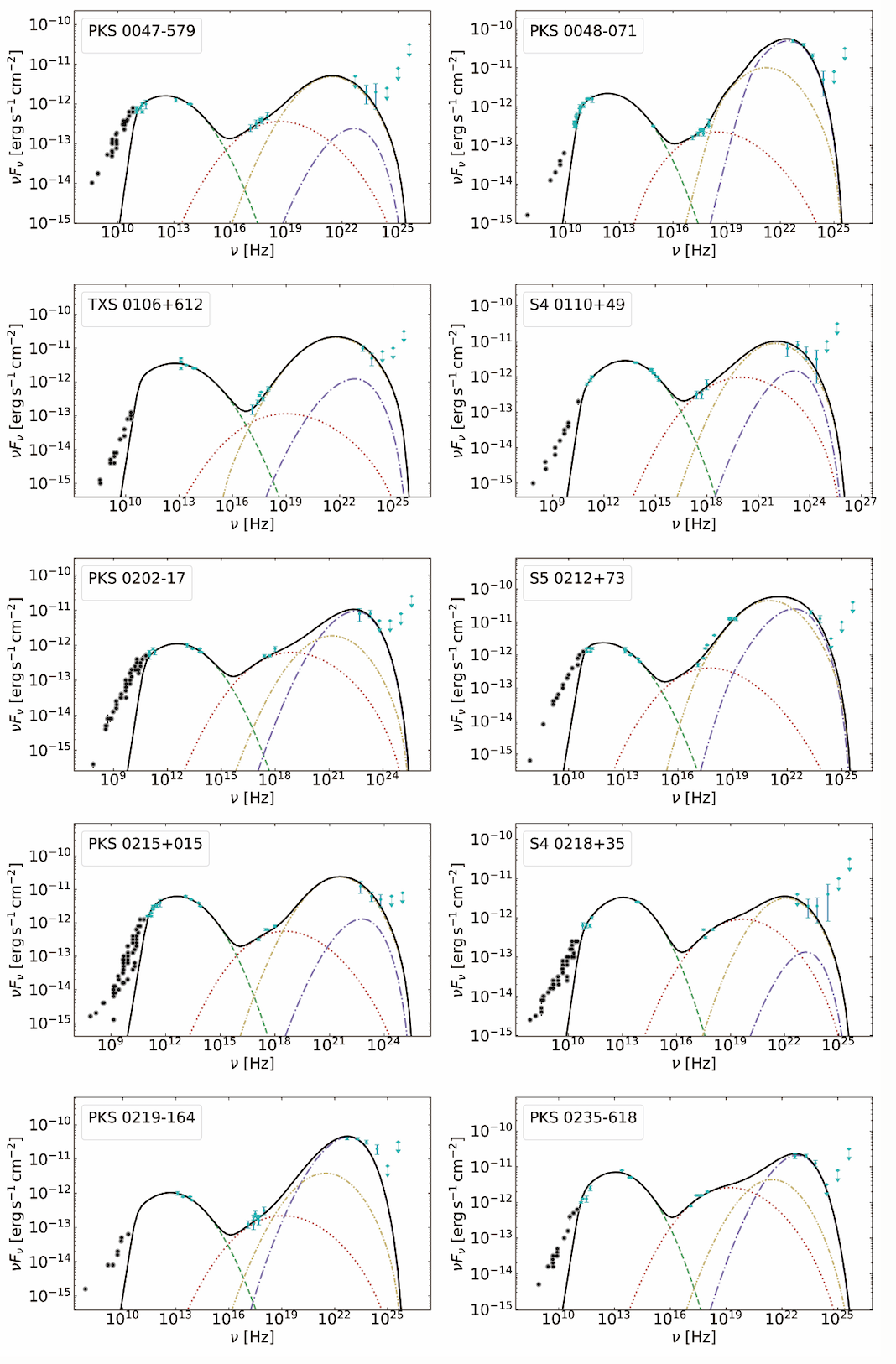}
 \caption{Comparisons of predicted multi-wavelength spectra with observed data for PKS 0047-579, PKS 0048-071, TXS 0106+612, S4 0110+49, PKS 0202-17, S5 0212+73, PKS 0215+015, S4 0218+35, PKS 0219-164, and PKS 0235-618, respectively. The radio data are shown in gray, and the quasi-simultaneous data from infrared to gamma-ray band are shown in blue. The green dashed line represents the synchrotron emission, the red dotted line represents the SSC emission, the purple dashed line and the yellow dashed line represents the EC emission, which seed photons from the BLR and DT, respectively, and the black solid curve is the total emission by summation of all the emission components.}
 \label{fig1:fig10}
\end{figure}

\begin{figure}{}
\centering
\includegraphics[width=1.7\columnwidth]{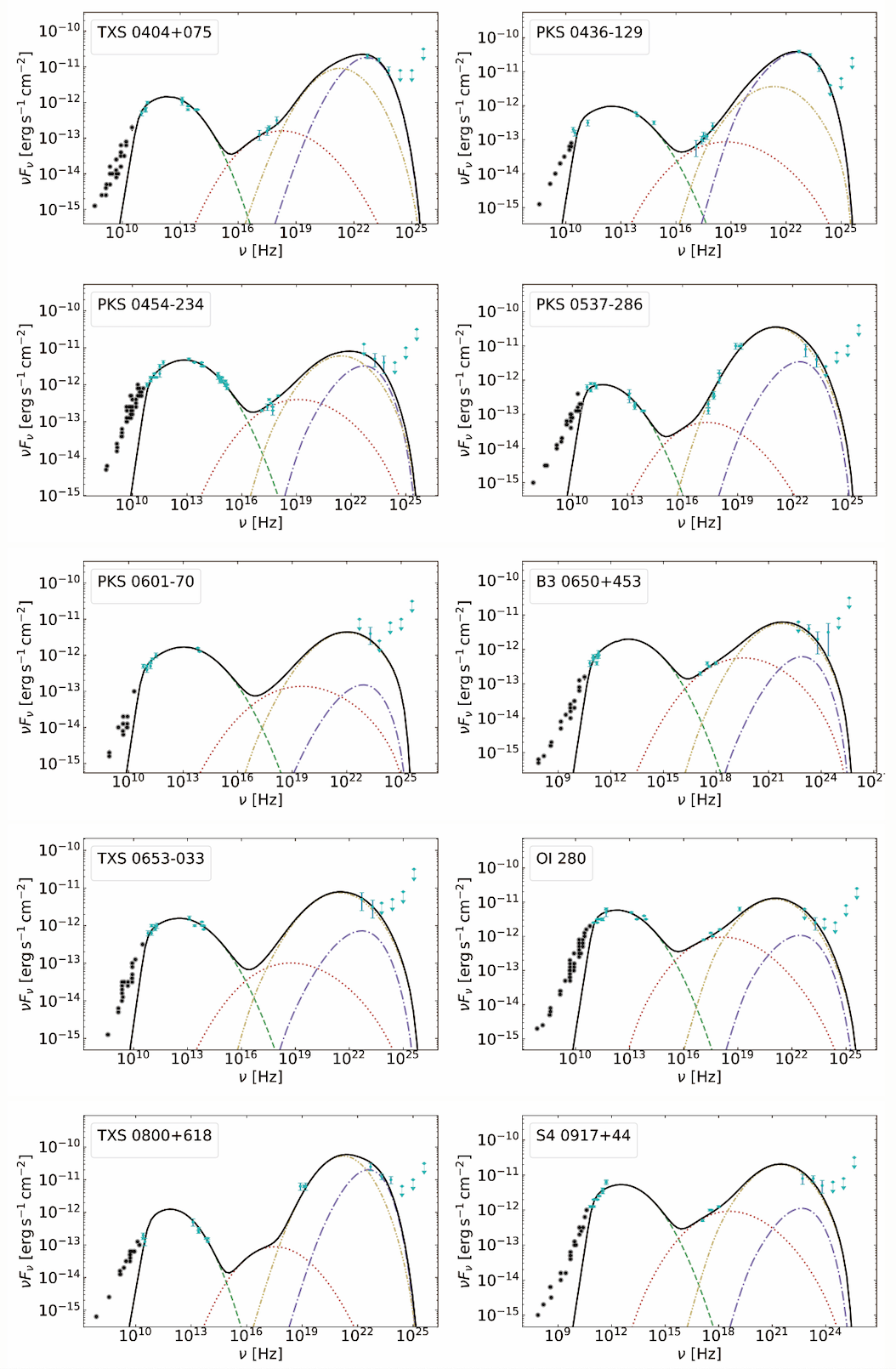}
 \caption{Comparisons of predicted multi-wavelength spectra with observed data for TXS 0404+075, PKS 0436-129, PKS 0454-234, PKS 0537-286, PKS 0601-70, B3 0650+453, TXS 0653-033, OI 280, TXS 0800+618, and S4 0917+44, respectively. The symbols and lines are the same as shown in Figure 1.}
 \label{fig11:fig20}
\end{figure}

\begin{figure}{}
\centering
\includegraphics[width=1.7\columnwidth]{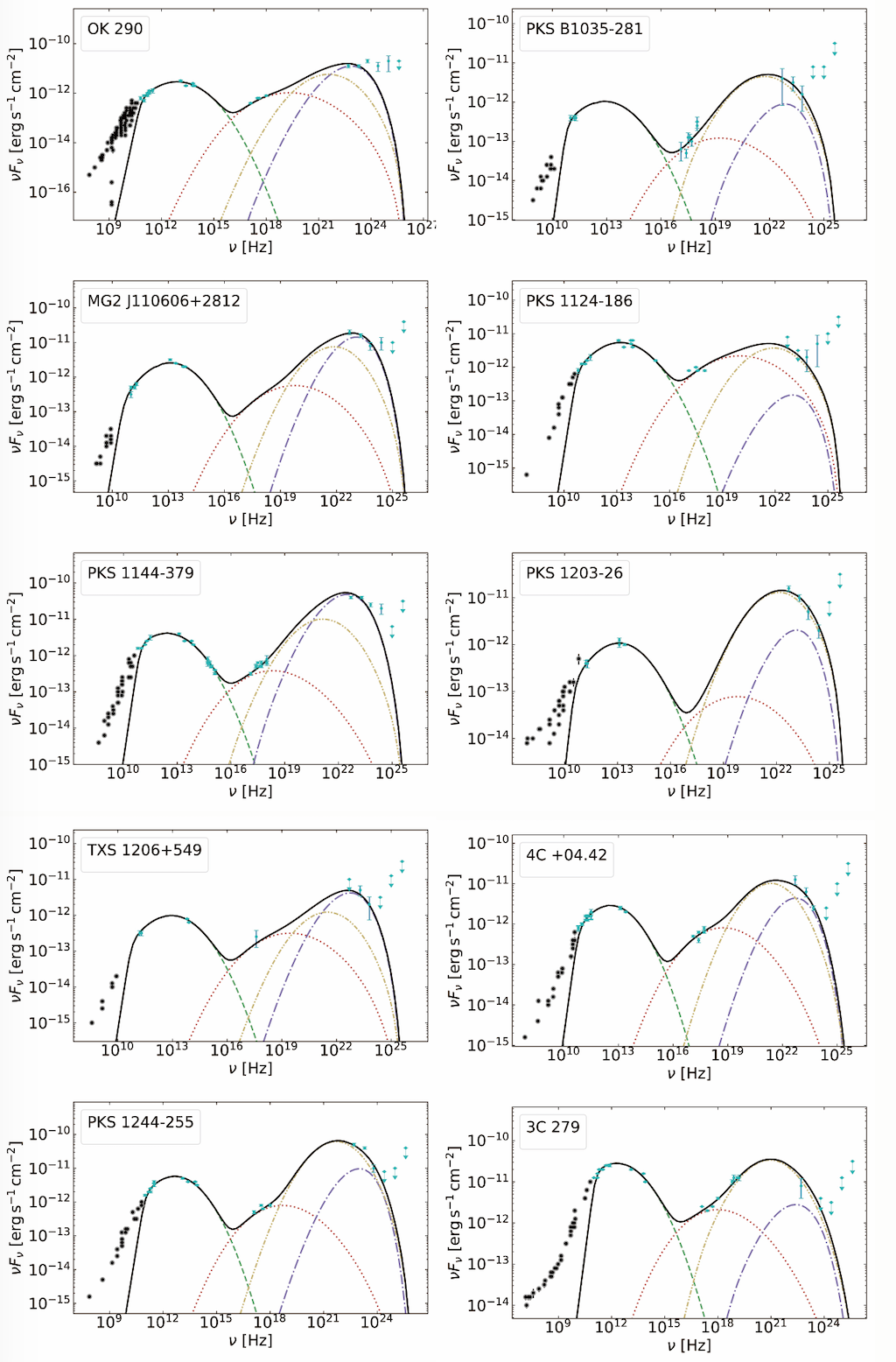}
\caption{Comparisons of predicted multi-wavelength spectra with observed data for OK 290, PKS B1035-281MG2 J110606+2812, PKS 1124-186, PKS 1144-379, PKS 1203-26, TXS 1206+549, 4C +04.42, PKS 1244-255, 3C 279, respectively. The symbols and lines are the same as shown in Figure 1.}
 \label{fig21:fig30}
\end{figure}

\begin{figure}{}
\centering
\includegraphics[width=1.7\columnwidth]{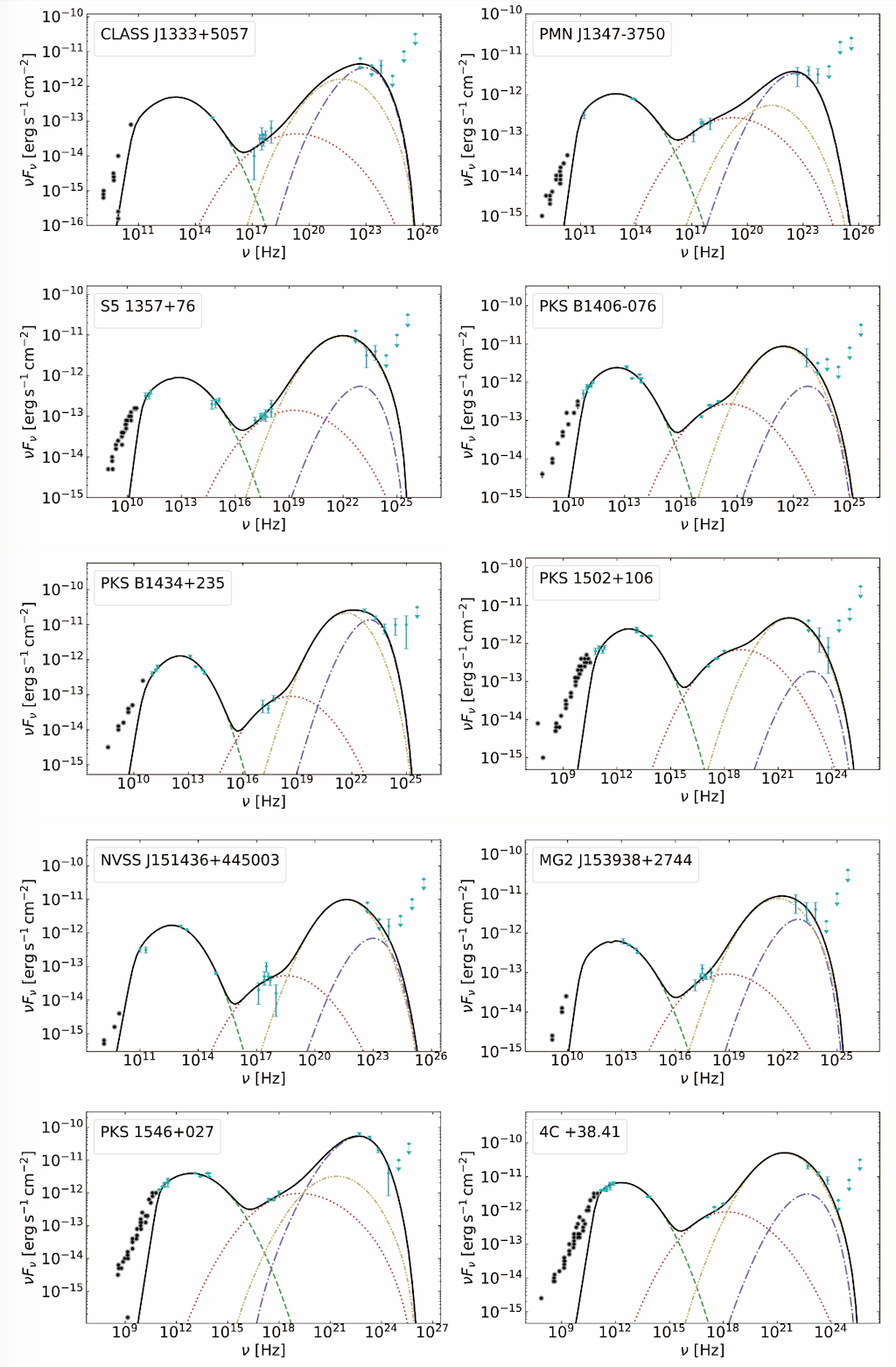}
 \caption{Comparisons of predicted multi-wavelength spectra with observed data for CLASS J1333+5057, PMN J1347-3750, S5 1357+76, PKS B1406-076, PKS B1434+235, PKS 1502+106, NVSS J151436+445003, MG2 J153938+2744, PKS 1546+027, 4C +38.41, respectively. The symbols and lines are the same as shown in Figure 1.}
 \label{fig31:fig40}
\end{figure}

\begin{figure}{}
\centering
\includegraphics[width=1.7\columnwidth]{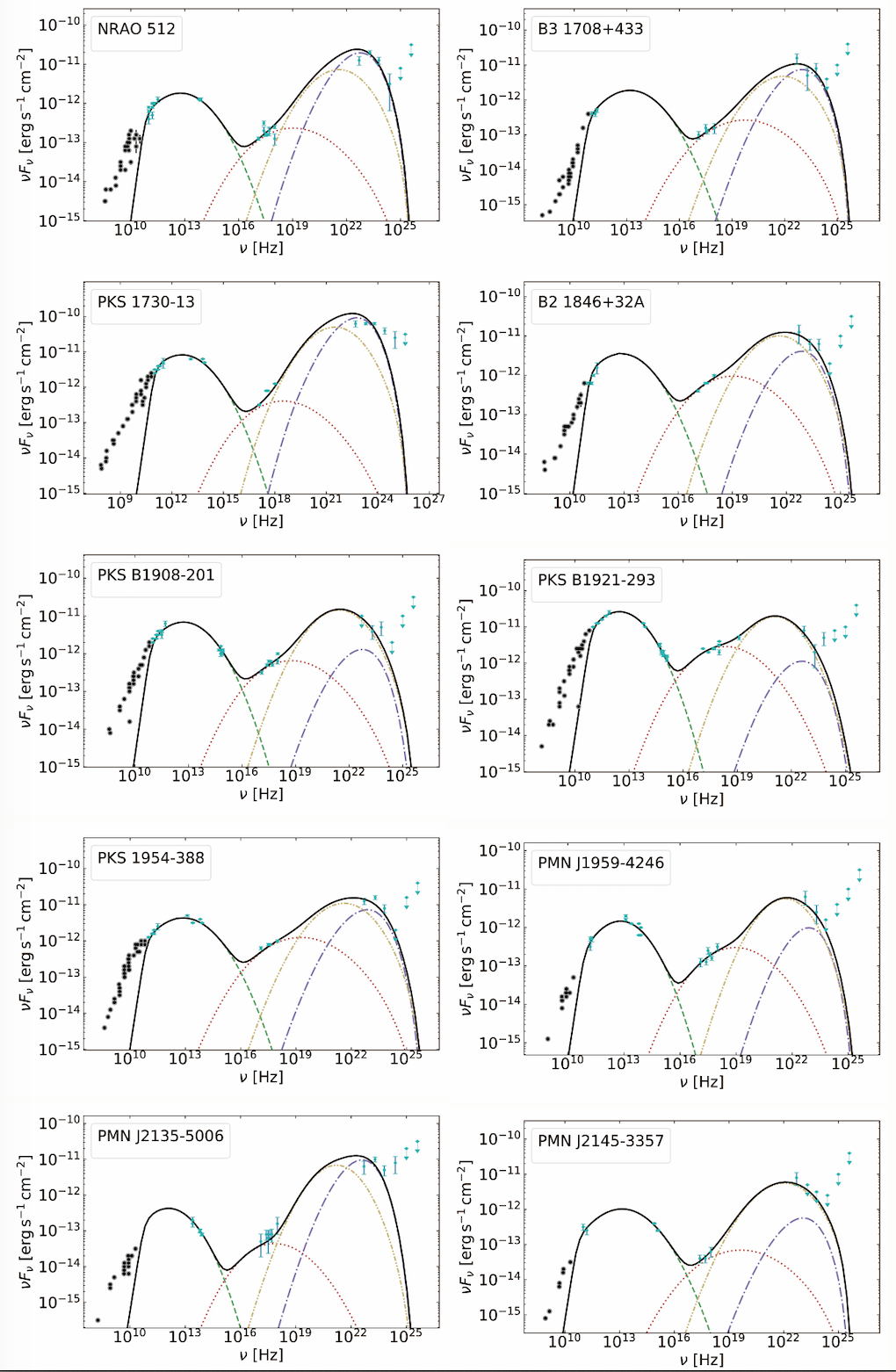}
 \caption{Comparisons of predicted multi-wavelength spectra with observed data for NRAO 512, B3 1708+433, PKS 1730-13, B2 1846+32A, PKS B1908-201, PKS B1921-293, PKS 1954-388, PMN J1959-4246, PMN J2135-5006, PMN J2145-3357, respectively. The symbols and lines are the same as shown in Figure 1.}
 \label{fig41:fig50}
\end{figure}

\begin{figure}{}
\centering
\includegraphics[width=1.7\columnwidth]{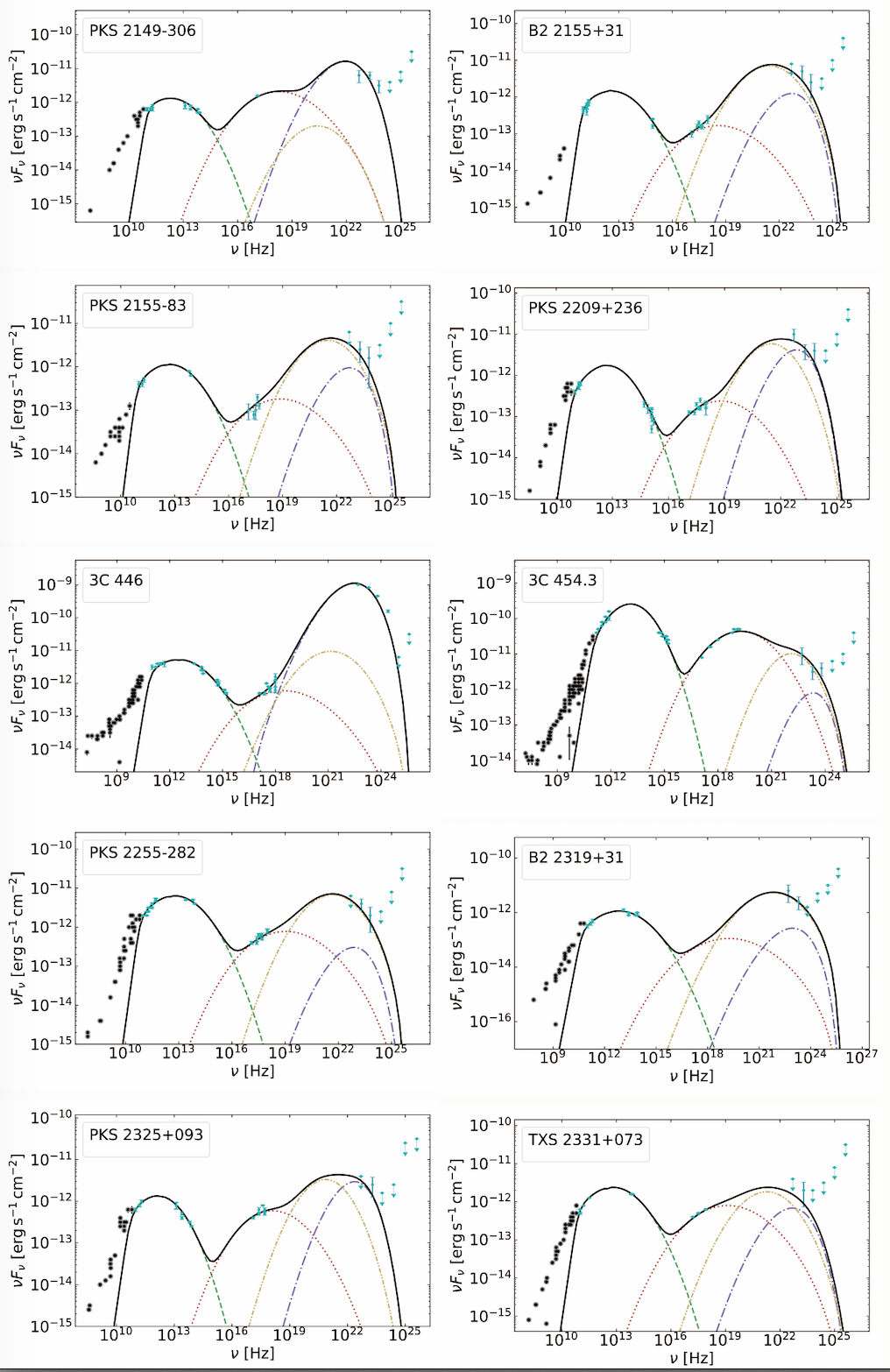}
 \caption{Comparisons of predicted multi-wavelength spectra with observed data for PKS 2149-306, B2 2155+31, PKS 2155-83, PKS 2209+236, 3C 446, 3C 454.3, PKS 2255-282, B2 2319+31, PKS 2325+093, TXS 2331+073, respectively. The symbols and lines are the same as shown in Figure 1.}
 \label{fig51:fig60}
\end{figure}

We apply our model to fit the SEDs of 60 $Fermi$ FSRQs, and the fitting results are shown in Figure~\ref{fig1:fig10} $-$ Figure~\ref{fig51:fig60}. In these Figures, the radio data are shown in black circles, and the quasi-simultaneous data from infrared to gamma-ray band are shown in cyan circles. The green dashed line represents the synchrotron emission, the red dotted line represents the SSC emission, the purple and the yellow dashed line represent the EC emission, in which seed photons are from the BLR and DT, respectively, and the black solid curve is the total emission by summarizing of all the emission.

When assuming that the multi-wavelength emission is from one dissipation region, the required number density of relativistic electrons is so high that the synchrotron emission below the turnover frequency (normally $\nu < 10^{11}~\rm Hz$) is inevitably self-absorbed. Previous studies suggest that it would be more natural to explain the GHz radio data by the synchrotron emission produced from extended jet \citep[e.g.,][]{2009MNRAS.399.2041G, 2010MNRAS.402..497G}. This work focuses on fitting the simultaneous broadband emission from inner jet, therefore we do not explain the radio data below the turnover frequency.

\section{Results and Discussion}\label{RD}

\subsection{The distributions}

\begin{figure}
\centering
\includegraphics[width=2\columnwidth]{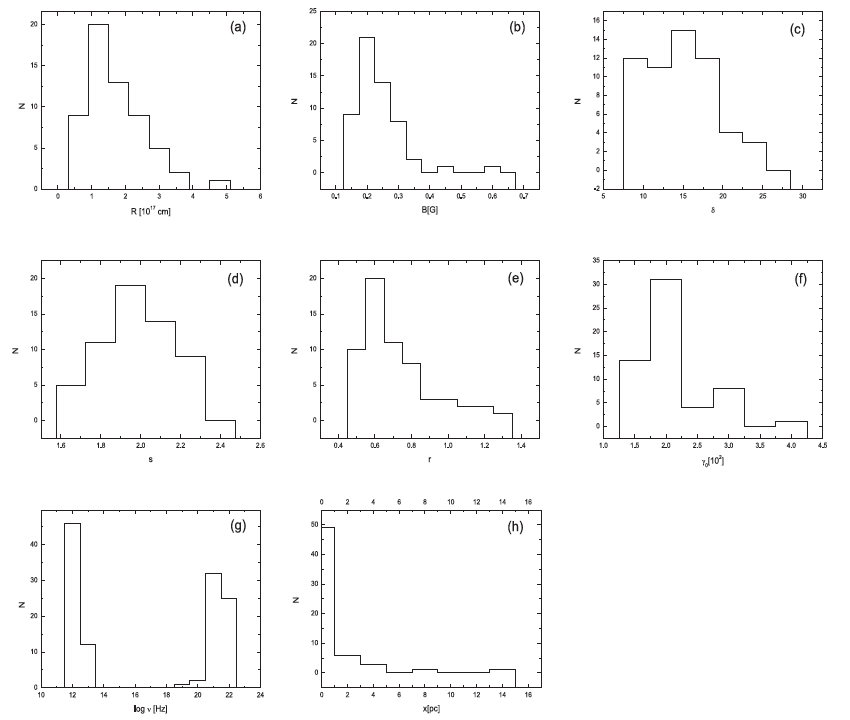}
 \caption{Distribution of the radiation region $R (\rm a)$, the magnetic field $B (\rm b)$, the Doppler factors $\delta (\rm c)$, the spectral index $s (\rm d)$, the spectral curvature $r (\rm e)$, the reference energy $\gamma_0 (\rm f)$, the two peak frequencies $\nu_{\rm syn}$ and $\nu_{\rm IC} (\rm h)$, and the distance $x (\rm h)$ between the dissipation region and the central black hole.}
\label{fig:7}
\end{figure}

In Figure~\ref{fig:7}(a)$-$(f), we show the parameter distributions.
%we give the distributions of the radiation region $R(\rm a)$, the magnetic field $B(\rm b)$, the doppler factors $\delta (\rm c)$, the spectral index $s(\rm d)$, the spectral curvature $r(\rm e)$, the reference energy $\gamma_0 (\rm f)$, the two peak frequencies $\nu_{\rm syn}$ and $\nu_{\rm IC}(\rm g)$, and the distance $x (\rm h)$ between the dissipation region and the central black hole.
The values of $R$ are in the range of (0.55 $\sim$ 4.95) $\times10^{17}$ cm.
The values of $B$ are in the range of 0.11 $\sim$ 0.62 G, which is similar to \citet{1998MNRAS.301..451G}, \citet{2008MNRAS.385..283C}, \citet{2011MNRAS.414.2674G}. The values of $\delta$ are in the range of 7 $\sim$ 27, which is consistent with \citet{1998MNRAS.301..451G} and observations \citep{2009A&A...494..527H}.
The range of $s$ are from 1.57 $\sim$ 2.4, and the range of $r$ are from 0.5 $\sim$ 1.3. % We calculate the synchrotron curvature using the relationship of $b_{sy} = r/5$ of \citet{2014ApJ...788..179C}, and the $b_{sy}$ varies between 0.1 and 0.262.
We also show the distributions of the synchrotron peak frequency $\nu_{\rm s}$ and the IC peak frequency $\nu_{\rm IC}$ in Figure~\ref{fig:7}(g), the values of $\nu_{\rm s}$ are in the range of $6.76\times10^{12}\sim 1.81\times10^{13}$ Hz and the values of $\nu_{\rm IC}$ are in the range of $8.47\times10^{20}\sim 5.62\times10^{22}$ Hz. Our parameters are all within a reasonable range.

\subsection{Particle acceleration mechanisms}

There are two mechanisms, statistical and stochastic acceleration mechanisms, can produce the electron energy distribution that follows the log-parabolic law, resulting in a log-parabolic SED. In the framework of statistical acceleration which needs an energy-dependent acceleration probability, the electron energy distribution follows the log-parabolic law when acceleration efficiency is inversely proportional to their energy \citep{2004A&A...413..489M}. In this process, \citet{2014ApJ...788..179C} suggested that the correlation between $\nu_{\rm p}$ and $b_{\rm syn}$ is $\log \nu_{\rm p}\approx 2/(5b_{\rm syn})+m$. While, In the framework of statistical acceleration which needs a fluctuations of fractional acceleration gain, the electron energy distribution follows the log-normal law when the energy gain fluctuations are a random variable around the systematic energy gain \citep{2011ApJ...739...66T}. In this process, $\nu_{\rm p}$ and $b_{\rm syn}$ follow the correlation of $\log \nu_{\rm p}\approx 3/(10b_{\rm syn})+m$ \citep{2014ApJ...788..179C}. In the framework of stochastic acceleration, the log-parabolic distribution can be derived from a mono-energetic and instantaneous injection \citep{2011ApJ...739...66T}. In this process, \citet{2014ApJ...788..179C} obtained the correlation of $\log \nu_{\rm p}\approx 1/(2b_{\rm syn})+m$. Therefore, the theoretical expected values of slope are 10/3, 5/2 and 2 for the fractional acceleration gain fluctuation, energy-dependent acceleration probability and stochastic acceleration processes, respectively.

\begin{figure}
\epsscale{2.0} \plotone{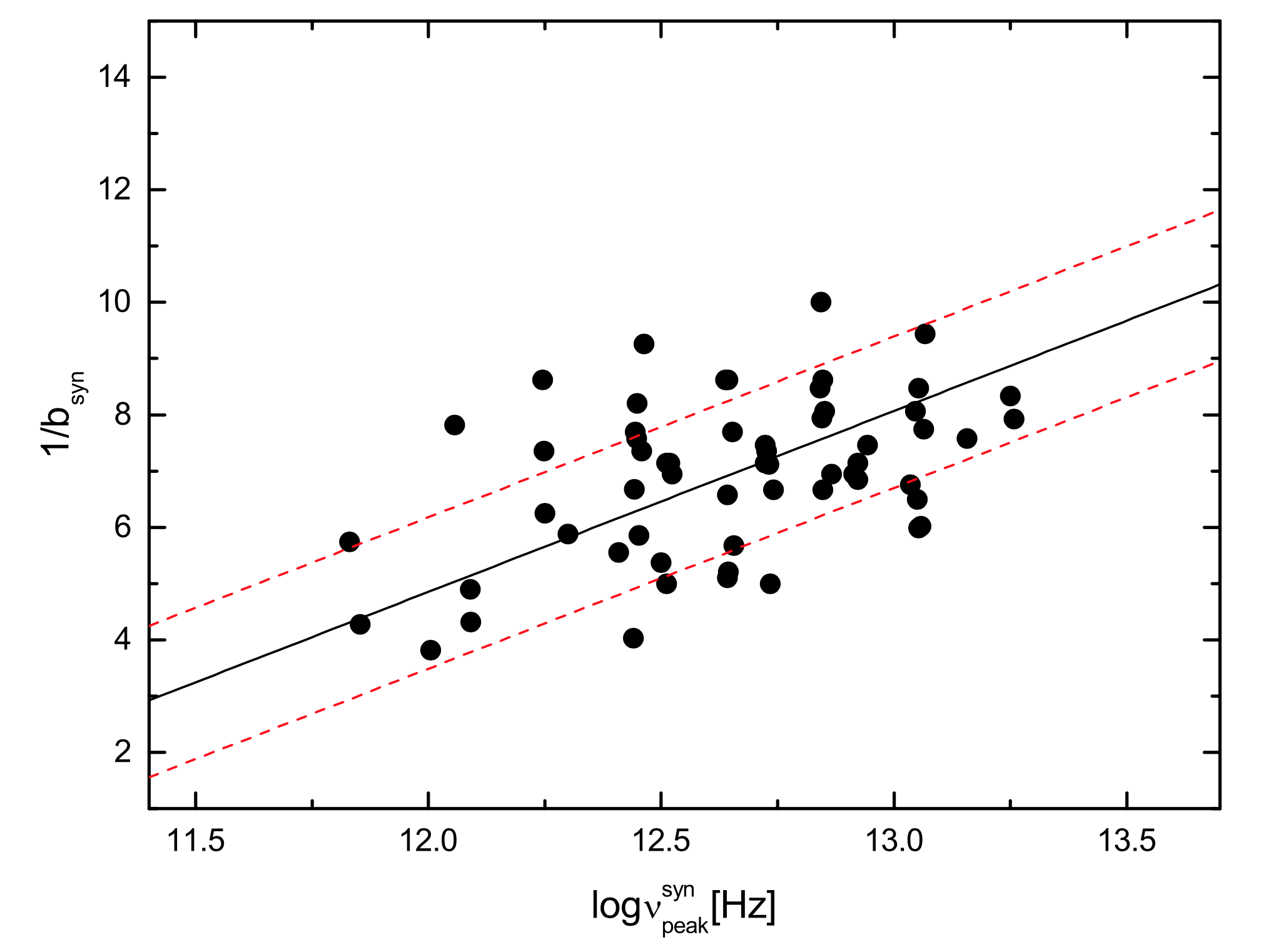}
\caption{The synchrotron peak frequency versus synchrotron curvature (in $1/b_{syn}$). The solid line is the best linear fit($p=2.958\times 10^{-4}$) and the dashed red lines indicate 1$\sigma$ confidence bands.}
\label{fig:8}
\end{figure}

In order to study the particle acceleration of our sample, we plot the synchrotron peak frequency versus its curvature in Figure~\ref{fig:8}. In order to facilitate comparison with these theoretical values of \citet{2014ApJ...788..179C}, we use $1/b_{\rm syn}$ ($b_{\rm syn}=r/4$) to represent the synchrotron curvature. The Spearman test gives a significance level $p=2.958\times 10^{-4}$ and a coefficient of correlation $\rho=0.464$. The bisector linear regression gives the best linear fitting equation as $1/b_{\rm syn}=(3.22\pm 0.52)\rm log\nu_{\rm p}-(33.74\pm0.52)$, represented as a solid black line in Figure~\ref{fig:8}. The dashed red lines indicate the 1$\sigma$ confidence bands. We find that the frequency of the synchrotron peak is moderately correlated to its curvature because of the following two reasons. Firstly, it may be caused by the existence of different acceleration mechanisms for each FSRQ. Secondly, the synchrotron peak frequency $\nu_{\rm p}$ is estimated by 
\begin{equation}
\log{\gamma_{\rm peak}}=\log{\gamma_0}+(3-s)/2r,
\end{equation}
\begin{equation}
\nu_{\rm p}=3.7\times10^6\gamma_{\rm peak}^2B\delta/(1+z).
\end{equation}
It can be seen that $\nu_{\rm p}$ is derived by six free parameters that vary from source to source and all can contribute to the scatter of the $b_{\rm syn}-\nu_{\rm p}$ correlation. Therefore, it is natural to find that the significance of this correlation will be weakened. In spite of this, for our sample, the slope is $k=3.22\pm 0.52$, which is basically consistent with statistical acceleration in a fluctuation of fractional acceleration gain \citep{2014ApJ...788..179C}.

\citet{2014ApJ...788..179C} analyzed the correlation of $\nu_{\rm p}$ and $b_{\rm syn}$ by fitting the broadband SEDs of 43 blazars with the second-degree polynomial function. He found the slope of $\rm log\nu_{\rm p}-1/b_{\rm syn}$ relation was $2.04\pm0.03$, and suggested that the stochastic acceleration mechanism is dominant in jet. However his sample is too small to separate them into FSRQs and BL Lacs. Moreover, FSRQs and BL Lacs are two different subclasses of AGN, so they may have different acceleration mechanisms. \citet{2016MNRAS.463.3038X} collected a much larger sample which is separated into FSRQs and BL Lacs, and found they show different correlation. For BL Lacs, the slop ($k=1.87\pm0.19$) is consistent with the stochastic acceleration mechanism. For FSRQs, the slope ($k=3.69\pm024$) is closely consistent with the statistic acceleration for the case of fluctuation of fractional acceleration gain. This particle acceleration mechanism of FSRQs is consistent with our modeling result. For some individual sources, \citet{2016ApJ...831..102K,2017ApJ...848..103K,2018ApJ...858...68K} analysed the Swift-XRT observations of Mrk 421 in different periods, and found that the correlation of $E_{\rm p}- b$ is weak or very weak. In addition, \citet{2019ApJ...885....8W} studied the $E_{\rm p}-b$ relation of 14 BL Lac objects using the log-parabolic model \citep{2004A&A...413..489M}, and found no correlation between $E_{\rm p}$ and $b$. The possible reasons are as follows: (1) They used the Swift-XRT data points which are not enough to distribute a complete synchrotron peak spectrum, resulting in a different curvature values and no correlation between $E_{\rm p}$ and $b$. (2) For a single source \citep[like Mrk 421 in][]{2016ApJ...831..102K,2017ApJ...848..103K,2018ApJ...858...68K}, there may be several acceleration mechanisms (first-order Fermi acceleration, magnetic reconnection, shear acceleration, and stochastic acceleration etc.) and they compete with each other \citep{2018ApJ...858...68K}, resulting in a weakness or even absence of $E_{\rm p}$ and $b$ correlation. For a sample that contains more than tens objects, there will be a tendency for only one acceleration mechanism to dominate. For our sample, the dominant acceleration mechanism is statistical acceleration with a fractional acceleration gain.

\subsection{Location of $\gamma$-ray emission region}
\begin{figure}
\epsscale{2.0} \plotone{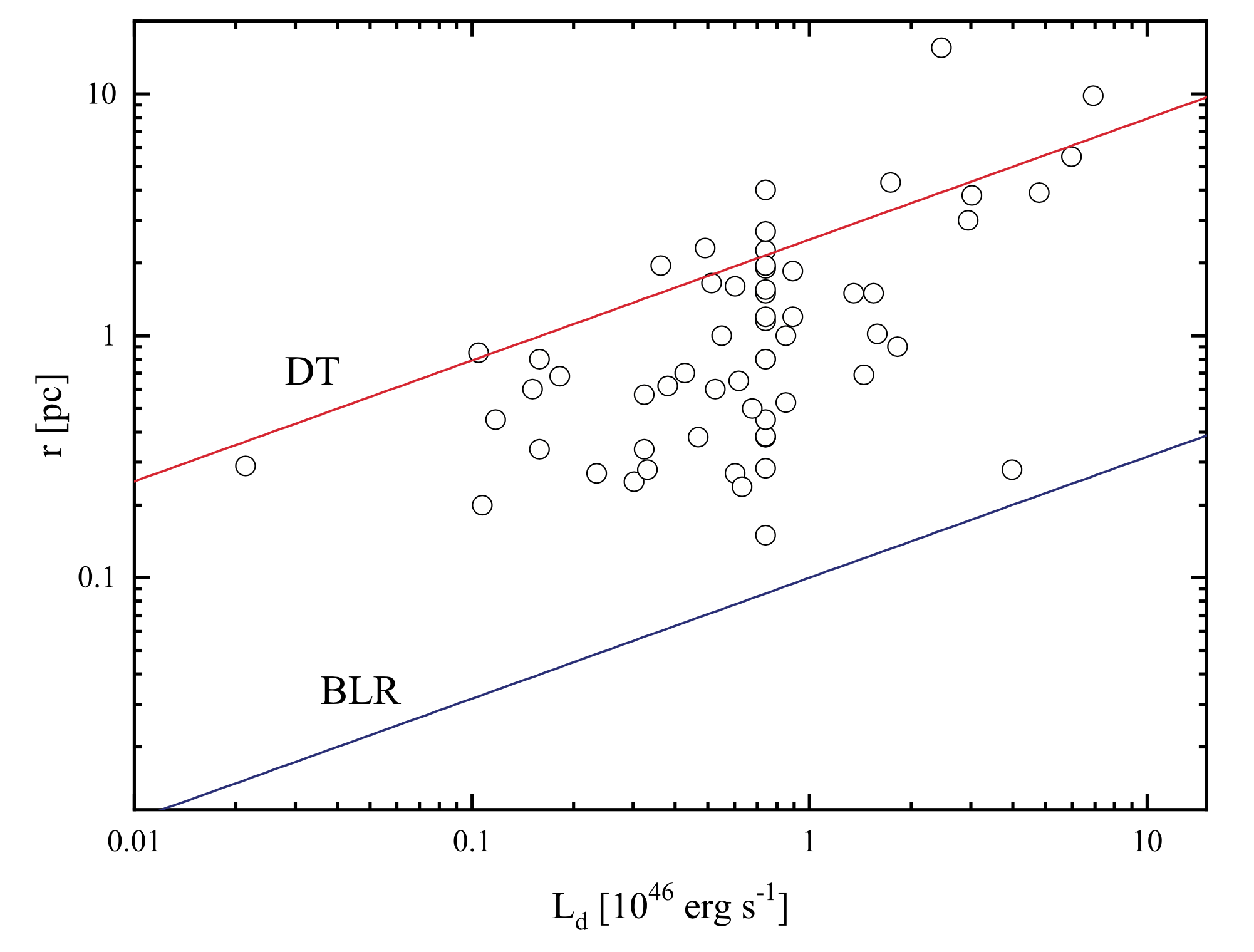}
\caption{The location of the $\gamma$-ray emission region from the SMBH ($x$) as a function of the luminosity of an accretion disk ($L_d$). The sample sources are exhibited  in open circles. The blue and red solid show the distances of BLR and DT, respectively.}
\label{fig:9}
\end{figure}

The location of the $\gamma$-ray emitting region in blazars is still controversial. In general, the energy dissipation region is constrained by two diagnostics, which are finding the variability timescale \citep[e.g.,][]{2010ApJ...712..957A,2011MNRAS.418...90L,2012ApJ...751L...3G,2013MNRAS.431..824B,2015MNRAS.452.1280R} and fitting the SED \citep[e.g.,][]{2009ApJ...692...32D,2012MNRAS.421.2956Z,2014ApJS..215....5K,2017ApJS..228....1Z,2018ApJ...859..168Y}. In our work, we fit the quasi-simultaneous SEDs of 60 FSRQs with the one-zone leptonic model to constrain it. Here, we assume that the energy density of the external ambient fields are functions of the distance between the position of the dissipation region ($x$) and SMBH (see Equations.~\ref{eq:quadratic6} and ~\ref{eq:quadratic7}), where the $x$ is determined by reproducing the $\gamma$-ray spectra. In Figure~\ref{fig:9}, we plot $x$ as a function of the luminosity of an accretion disk ($L_d$) for our sample. It can be seen that the $\gamma$-ray dissipation regions in our modeling are located at the range from 0.1 to 10 pc, which means that most of them are located outside the BLR and within the DT. In the modeling, the soft photons for the EC radiation of 40 FSRQs are dominated by that from DT and the soft photons of other 20 FSRQs are dominated by that from BLR. Therefore, we find that the gamma-ray emission regions are located outside the BLR, and the soft photons for EC processes of most FSRQs are dominated by that of DT.%According to the flow formulas of DT and BLR, it is found that the BLR is more advantageous than the DT. However, the EC process of BLR is prone to truncation of the energy spectrum due to the Klein-Nishina effect, so the DT is used to fit for most sources.

Our results are consistent with many previous works. For example, \citet{2014MNRAS.441.1899F} used the `time lag core shift' method for quasar 3C 454.3 to estimate a lower limit for the distance of the bulk $\gamma$-ray production region and suggested that the gamma-ray emission region are $\sim 0.8$-$1.6$ pc away from the SMBH. \citet{2018arXiv180904984J} used the lag times of PMN J2345-1555 to derive the optical and the $\gamma$-ray emitting regions coincide, which are located at $4.26 ^{+0.83} _ {-0.79}$ pc away from 15 GHz core position in jet. %\citet{2012arXiv1202.6193G} suggested that $\gamma$-ray emission region is closer to the DT range than that of the BLR through their proposed diagnostic method: the connection of three observable values, synchrotron and external Compton peak frequencies and the Compton dominance, to the seed photon energy and energy density.
In addition, if the gamma-ray dissipation regions are located inside BLR, the $\gamma$ rays are likely to be absorbed through the $\gamma\gamma$ pair production. However, only $10\%$ of blazars show a feature of attenuation in their $\gamma$-ray spectrum \citep{2018MNRAS.477.4749C}. So, $\gamma$-ray should be produced in a region far from the SMBH.
% and our results are reliable if the VHE $\gamma$-ray emission can be observed from FSRQs. There are many source have been observed the VHE $\gamma$-ray emission. Such as 3C 279 \citep{2011A&A...530A...4A}, PKS1222+216 \citep{2011ApJ...730L...8A}, PKS 1510-089 \citep{2011ATel.3509....1H}, PKS 1441+25 \citep{2015ApJ...815L..22A}, and S30218+35 \citep{2015ICRC...34..825S}.
There are some different views. For example, \citet{2010ApJ...717L.118P} fitted the SED of 3C 454.3 using the broken power-law distribution, and suggested that the $\gamma$-ray emission region lies inside the region of the BLR. However, the broken power-law distribution of the $\gamma$-ray spectrum of 3C 454.3 could arise equally well from a break in the electron energy distribution. In addition, \citet{2015PASJ...67...79L} fitted the quasi-simultaneous SED of 4C +21.35 with a one-zone leptonic model. They suggested that the emitting regions locate within the BLR clouds and around the outer radius of the BLR during flaring states, while the emitting region locates beyond the DT, $x > x_{\rm DT} \simeq 4.8 $ pc, for quiescent state. Their results suggested that the location of the gamma-ray dissipation region is distinct in different states of source. Since the quasi-simultaneous data of our sample are in the steady state and the location of the emission regions are relatively diffuse for our sample, our results are more inclined to suggest that the dissipation zone is outside the BLR.

\subsection{The physical properties of jet}
The jet power $P_{\rm jet}$ is critical to understand the production and composition of the jets and we can estimate them through fitting the SED. We assume that the jet power is carried by relativistic electron, cold proton, magnetic field, and radiation, i.e.,
\begin{equation}
    P_{\rm jet}=\sum_i \pi R^2{\Gamma^2}c{U_i},
    \label{eq:quadratic10}
\end{equation}
where $U_i(i=\rm e,p,B,r)$ are the energy densities associated with the emitting electron $U_{\rm e}$, cold proton $U_{\rm p}$, magnetic field $U_{\rm B}$, and radiation $U_{\rm r}$ measured in the comoving frame \citep{2010MNRAS.402..497G}, which are given by
\begin{equation}
    U_{\rm e}=m_ec^2\int {N(\gamma)\gamma d\gamma},
    \label{eq:quadratic11}
\end{equation}

\begin{equation}
    U_{\rm p}=m_pc^2\int {N(\gamma)d\gamma},
    \label{eq:quadratic12}
\end{equation}

\begin{equation}
    U_{\rm B}=B^2/{8\pi},
    \label{eq:quadratic13}
\end{equation}

\begin{equation}
    U_{\rm r}=\frac{L_{\rm obs}}{4\pi R^2c\delta^4},
    \label{eq:quadratic14}
\end{equation}
where $L_{\rm obs}$ is the total observed non-thermal luminosity, which is calculated from the model. Here, we assume that the charge neutrality is provided by one cold proton per relativistic electron (i.e. no electron-positron pairs). The calculated $P_{\rm e}$, $P_{\rm p}$, $P_{\rm B}$, $P_{\rm r}$, and $P_{\rm jet}$ are listed in Table~\ref{table3}.

\begin{figure}
\epsscale{2.0} \plotone{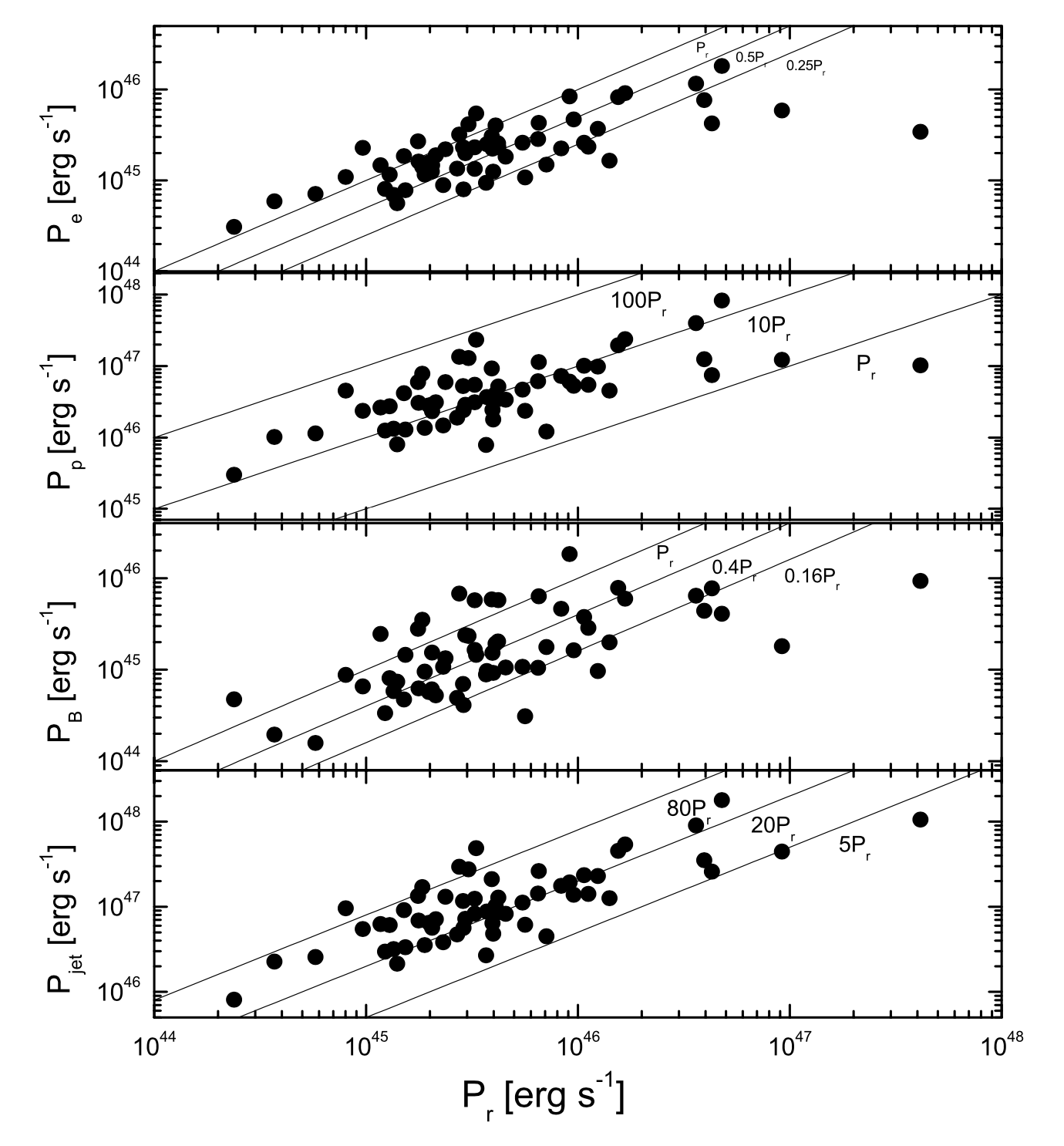}
\caption{The $P_{\rm e},P_{\rm p},P_{\rm B},P_{\rm jet}$ as a function of the $P_{\rm r}$.}
\label{fig:10}
\end{figure}

In Figure~\ref{fig:10}, we plot $P_{\rm e}$, $P_{\rm p}$, $P_{\rm B}$, and $P_{\rm jet}$ as a function of $P_{\rm r}$. The relation $P_{\rm e}\sim P_{\rm B}\leq P_{\rm r}$ $<$ $P_{\rm p}$ can be found. The $P_{\rm e}\sim P_{\rm B}$ suggests that SEDs fitted with parameters that are close to the condition of equipartition between the magnetic field and the relativistic electrons, which is consistent with many previous studies, such as \citet{2013ApJ...768...54B}. Both of $P_{\rm e}$ and $P_{\rm B}$ are in the range of $10^{44}-10^{46} \rm erg~s^{-1}$. For example, the $P_{\rm e}$, $P_{\rm B}$ and $P_{\rm p}$ of 3C 279 and 3C 454.3 obtained also similar to those of \citet{2013ApJ...768...54B}, indicating that the radiation processes and radiation efficiencies in the jet are similar even if they are modeled with different parameters and electron energy distributions. They also used the hadronic model to fit the SEDs of blazars. However, the characteristic break of two quasars of their sample cannot be well modeled at a few GeV energies, and the modeling required powers in relativistic protons are in the range $P_{\rm p}$ $\sim$ $10^{47}-10^{49} \rm erg~s^{-1}$, which exceed their Eddington luminosity. The $P_{\rm B}\leq P_{\rm r}$ implies that Poynting flux cannot account for the radiation power. It can be seen that $P_{\rm r}/P_{\rm e} \sim 0.5 - 4$, which indicates that a large fraction of the relativistic electron power would be used to produce the observed radiation. The result $P_{\rm e} < P_{\rm r}$ derived here suggest that the most energy of the relativistic electrons are dissipated by EC radiation.

\begin{figure}
\epsscale{2.0} \plotone{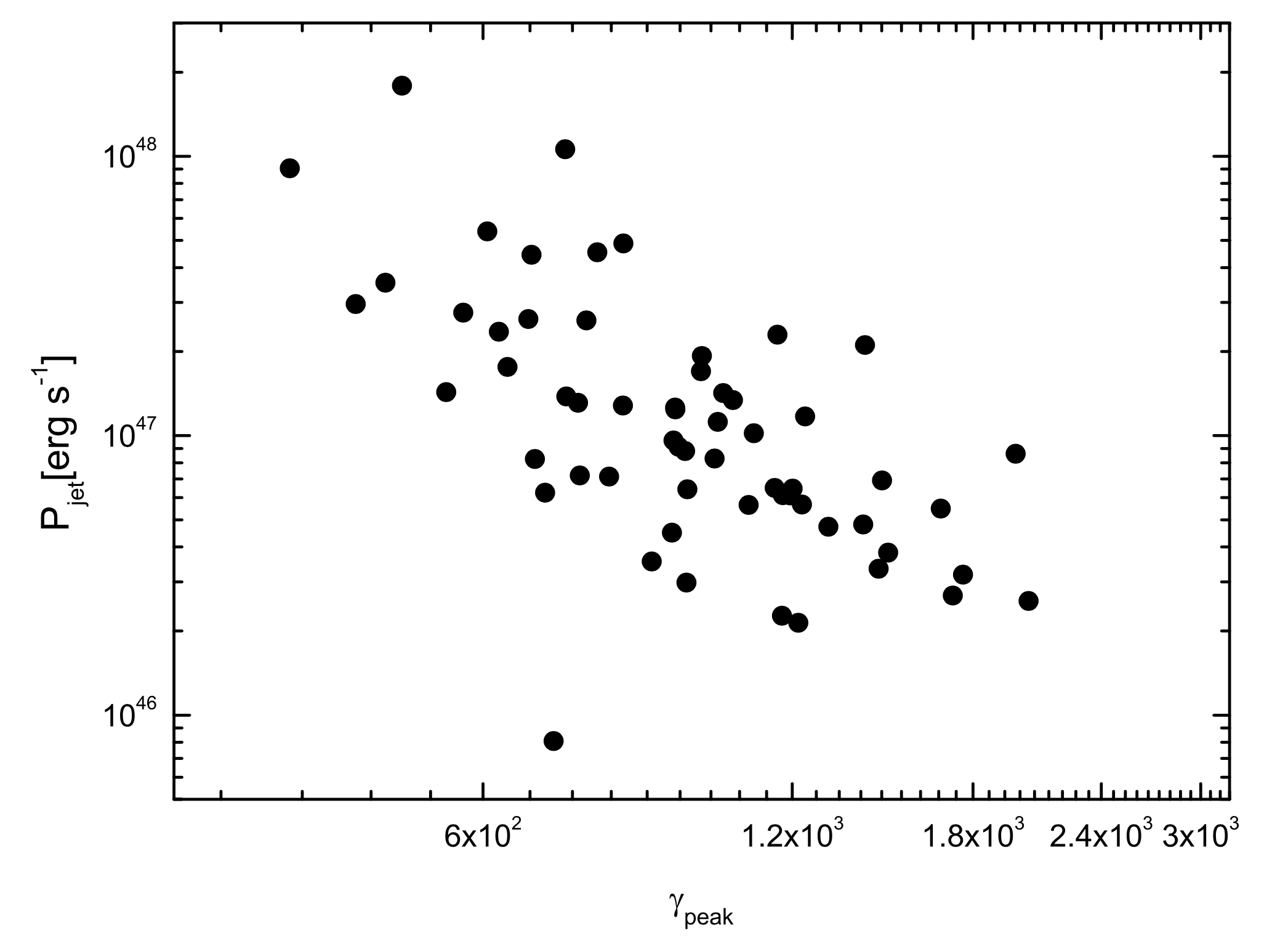}
\caption{The $P_{\rm jet}$ as a function of the $\gamma_{\rm peak}$.}
\label{fig:11}
\end{figure}

There is an anti-correlation between $\gamma_{\rm peak}$ and $P_{\rm jet}$ in our sample, i.e, $P_{\rm jet} = -2.58 \gamma_{\rm peak} + 54.695$ with $\rho = -0.625$ and $p = 9.26 \times 10^{-8}$ (in Figure~\ref{fig:11}), where $\log{\gamma_{\rm peak}}=\log{\gamma_0}+(3-s)/2r$ is the peak energy of SED. This result is consistent with the prediction of the blazar sequence \citep{1998MNRAS.299..433F,1998MNRAS.301..451G,2008MNRAS.387.1669G,2008MNRAS.385..283C}, which is usually explained as that the radiation cooling is stronger in powerful blazar. \citet{2018PASJ...70....5Q} also found that the peak luminosity $L_{\rm peak}$ was inversely related to the peak frequency $\nu_{\rm peak}$ and $\gamma_{\rm peak}$ to $P_{\rm jet}$, both of which supported the blazar sequence. Although some works suggested that the sequence is the result of a selection effect \citep[e.g.,][]{2003ApJ...588..128P,2005A&A...434..385G}, many works proposed that the blazar sequence is still theoretically valid \citep[e.g.,][]{1998MNRAS.301..451G,2002ApJ...564...86B,2013ApJ...763..134F}. %\citet{2013ApJ...763..134F} suggested that the compton dominance is a more intrinsic indicator for blazar sequence. In addition, \citet{2017MNRAS.469..255G} discovered that FSRQs do form a sequence in Compton dominance and in the $\rm X$-ray slope. Comparing synchrotron peak and IC peak of each source from Figure~\ref{fig1:fig10} $-$ Figure~\ref{fig51:fig60}, it can be seen that the Compton are greater than the synchrotron peak energy flux, and these sources are Compton dominated.

\section{Conclusions}\label{con}
On the basis of the systematic SED fitting with the one-zone leptonic model for the quasi-simultaneous SEDs of 60 $Fermi$-4$\rm LAC$ FSRQs, we investigate the jet physical properties of $Fermi$-4$\rm LAC$ FSRQs. Our main results are summarized below.

\begin{itemize}
  \item [1.]
  We find the correlation between the curvature $1/b_{\rm syn}$ and peak frequency $\nu_{\rm p}$ is $1/b_{\rm syn}=(3.22\pm 0.52)\rm log\nu_{\rm p}-(33.74\pm0.52)$, and the slope $k=3.22\pm 0.52$ is consistent with statistical acceleration in a fluctuation of fractional acceleration gain. We suggest that statistical acceleration with a fractional acceleration gain is the dominant acceleration mechanism for FSRQs in the steady state SEDs .
 \item [2.]
  The gamma-ray dissipation region is distinct in different states of source. We find that the $\gamma$-ray dissipation regions for most FSRQs in the steady state are located at the range from 0.1 to 10 pc during quiescent state, which mean that they are located outside the BLR and within the DT.
\item [3.]
  We find a size relation $P_{\rm e}\sim P_{\rm B}\leq P_{\rm r}$ $<$ $P_{\rm p}$. The $P_{\rm e}\sim P_{\rm B}$ suggests that SEDs fitted with parameters are close to equipartition between the magnetic field and the relativistic electrons. The $P_{\rm e} < P_{\rm r}$ suggest that the most energy of the relativistic electrons are dissipated by EC radiation for FSRQs. Then, we find an anti-correlation between $P_{\rm jet}$ and $\gamma_{\rm peak}$, which is consistent with the prediction of the blazar sequence.

\end{itemize}

We thank the anonymous referee for insightful comments and constructive suggestions. Part of this work is based on archival data, software or online services provided by the SPACE SCIENCE DATA CENTER (SSDC). This work is supported by the Joint Research Fund in Astronomy (Grant Nos 10978019, U1431123) under cooperative agreement between the National Natural Science Foundation of China (NSFC) and Chinese Academy of Sciences (CAS)and the Provincial Natural Science Foundation of Yunnan (Grant No 2019FB009).
%\newpage

\appendix
\section{Fermi-LAT analysis}
\textsl{Fermi}-LAT is a pair-production telescope with large effective area and large filed of view, covering in the energy band $20->300$ GeV, which provides a full-sky coverage every 3 hr \citep{2009ApJ...697.1071A}. We used the SOURCE class events converting in both the front and back sections, but rejecting that above a zenith angle of $90^\circ$ to reduce contamination by Earth limb emission. For each FSRQs, we selected the data collecting in the time interval ($\sim$ 2 months), in which observation from infrared to X-ray energy were performed simultaneously. Events extracted are within a $10^\circ$ region of interest(ROI) centered at the position of each FSRQs and with energies in the range between 100 MeV and 100 GeV.

 We employed the latest release \textsl{Fermi}-LAT Science Tool to determine the spectra in \textsl{Fermi}-LAT band. We derived the spectral points in 6 logarithmically spaced energy bins over 0.1-100 GeV for each target source. For each bin, we carried out an unbinned likelihood analysis modeling the target object as point sources with simple power-law photon spectra fixed index $\Gamma=2$. The background model comprise the diffuse components(i.e., Galactic diffuse emission modeled by $ gll\_iem\_v7.fits$ and isotropic emission with spectrum shape by $iso\_P8R3\_SOURCE\_V2\_v01.txt$), and the Fourth Source Catalog sources \citep[4FGL;][]{2020ApJS..247...33A} within each region of ROI enlarged by $5^\circ$ excluding our target object. For each energy bin fitting, we always left the normalization of the diffuse components free and fixed the spectral parameters of background 4FGL source to catalog value. Upper limits on the flux at $95\%$ confidence level were derived when the detection significance is lower than $2\sigma$ in a given bin.

%===================================
\newpage

\newpage

\newpage

\newpage

\newpage

\newpage
%===============================================================

%==============================
\newpage

%===================================
\newpage
\begin{table}
\scriptsize
\caption{Observation date between infrared to X-ray band, and the Integration date of $\gamma$ rays for the FSRQs in our sample\label{alltatatable}}
\tabcolsep 3mm
\begin{tabular}{ccccccc}
\hline\hline

Fermi Name & Source Name & RA(J2000) & Dec(J2000) & $\rm t_1^a$ & $t_2^b$ \\
~(1) & (2) & (3) & (4) & (5) & (6) \\
\hline
4FGL J0050.0-5736	&	PKS 0047-579	&	12.5197	&	-57.6164	&	2010/5/26-2010/6/1	&	2010/5/3-2010/7/3	\\
4FGL J0051.1-0648	&	PKS 0048-071	&	12.7824	&	-6.8096	&	2009/5/17	&	2009/4/17-2009/6/17	\\
4FGL J0109.7+6133	&	TXS 0106+612	&	17.445	&	61.5615	&	2010/1/31-2010/2/3	&	2010/1/1-2010/3/1	\\
4FGL J0113.4+4948	&	S4 0110+49	    &	18.3682	&	49.8054	&	2010/8/27	&	2010/7/27-2010/9/27	\\
4FGL J0205.0-1700	&	PKS 0202-17	    &	31.2637	&	-17.0022	&	2010/1/8	&	2009/12/8-2010/2/8	\\
4FGL J0217.4+7352	&	S5 0212+73	    &	34.3533	&	73.8804	&	2010/9/10-2010/9/11	&	2010/8/10-2010/10/10	\\
4FGL J0217.8+0144	&	PKS 0215+015	&	34.4621	&	1.7346	&	2010/1/29-2010/2/2	&	2010/1/1-2010/3/1	\\
4FGL J0221.1+3556	&	S4 0218+35	    &	35.281	&	35.9359	&	2010/8/4	&	2010/7/4-2010/9/4	\\
4FGL J0222.0-1616	&	PKS 0219-164	&	35.5197	&	-16.2787	&	2010/5/30	&	2010/4/30-2010/6/30	\\
4FGL J0236.8-6136	&	PKS 0235-618	&	39.2021	&	-61.6106	&	2010/6/12-2010/6/16	&	2010/5/12-2010/7/12	\\
4FGL J0407.5+0741	&	TXS 0404+075	&	61.8921	&	7.6998	&	2010/2/14-2010/2/15	&	2010/1/15-2010/3/15	\\
4FGL J0438.4-1254	&	PKS 0436-129	&	69.6099	&	-12.9053	&	2010/7/1	&	2010/6/1-2010/8/1	\\
4FGL J0457.0-2324	&	PKS 0454-234	&	74.2608	&	-23.4149	&	2010/2/24-2010/2/25	&	2010/1/25-2010/3/25	\\
4FGL J0539.9-2839	&	PKS 0537-286	&	84.9952	&	-28.6585	&	2010/3/7-2010/3/12	&	2010/2/7-2010/4/7	\\
4FGL J0601.1-7035	&	PKS 0601-70	    &	90.2958	&	-70.5895	&	2010/3/9	&	2010/2/8-2010/4/8	\\
4FGL J0654.4+4514	&	B3 0650+453	    &	103.606	&	45.2446	&	2010/3/23	&	2010/2/23-2010/4/23	\\
4FGL J0656.3-0322	&	TXS 0653-033	&	104.0765	&	-3.3686	&	2010/3/31	&	2010/3/1-2010/5/1	\\
4FGL J0750.8+1229	&	OI 280	        &	117.701	&	12.494	&	2010/4/5-2010/4/12	&	2010/3/8-2010/5/8	\\
4FGL J0805.4+6147	&	TXS 0800+618	&	121.3557	&	61.7937	&	2010/4/3-2010/4/4	&	2010/3/4-2010/5/4	\\
4FGL J0920.9+4441	&	S4 0917+44	    &	140.2291	&	44.699	&	2009/10/29	&	2010/9/29-2010/11/29	\\
4FGL J0956.7+2516	&	OK 290	        &	149.1761	&	25.2817	&	2010/5/7-2010/5/15	&	2010/4/25-2010/6/25	\\
4FGL J1037.7-2822	&	PKS B1035-281	&	159.4274	&	-28.3816	&	2010/1/22-2010/1/23	&	2009/12/22-2010/2/22	\\
4FGL J1106.0+2813	&	MG2 J110606+2812	&	166.502	&	28.2254	&	2010/5/24	&	2010/4/24-2010/6/24	\\
4FGL J1127.0-1857	&	 PKS 1124-186	&	171.7634	&	-18.964	&	2010/6/10	&	2010/5/10-2010/7/10	\\
4FGL J1147.0-3812	&	PKS 1144-379	&	176.76	&	-38.2006	&	2010/6/24	&	2010/5/24-2010/7/24	\\
4FGL J1205.7-2635	&	PKS 1203-26	    &	181.4322	&	-26.5946	&	2010/3/6-2010/3/7	&	2010/2/5-2010/4/5	\\
4FGL J1208.9+5441	&	TXS 1206+549	&	182.2261	&	54.6995	&	2010/5/16-2010/5/21	&	2010/4/18-2010/6/18	\\
4FGL J1222.5+0414	&	4C +04.42	    &	185.6271	&	4.2389	&	2010/6/17	&	2010/5/23-2010/7/23	\\
4FGL J1246.7-2548	&	PKS 1244-255	&	191.6887	&	-25.8018	&	2010/1/25	&	2009/12/25-2010/2/25	\\
4FGL J1256.1-0547	&	3C 279	        &	194.0415	&	-5.7887	&	2010/1/15	&	2009/12/15-2010/2/15	\\
4FGL J1333.7+5056	&	CLASS J1333+5057	&	203.4395	&	50.9366	&	2010/3/13	&	2010/2/13-2010/4/13	\\
4FGL J1347.6-3751	&	PMN J1347-3750	&	206.9135	&	-37.8633	&	2010/9/20	&	2010/8/20-2010/10/20	\\
4FGL J1358.1+7642	&	S5 1357+76	    &	209.5283	&	76.7064	&	2010/5/24-2010/5/25	&	2010/4/24-2010/6/24	\\
4FGL J1408.9-0751	&	PKS B1406-076	&	212.2356	&	-7.8575	&	2010/5/23	&	2010/4/23-2010/6/23	\\
4FGL J1436.9+2321	&	PKS B1434+235	&	219.2266	&	23.3611	&	2010/6/14	&	2010/5/14-2010/7/14	\\
4FGL J1504.4+1029	&	PKS 1502+106	&	226.1033	&	10.4978	&	2010/7/29	&	2010/6/29-2010/8/29	\\
4FGL J1514.8+4448	&	NVSS J151436+445003	&	228.7193	&	44.8105	&	2010/4/6	&	2010/3/6-2010/5/6	\\
4FGL J1539.6+2743	&	MG2 J153938+2744	&	234.9019	&	27.7277	&	2010/3/17	&	2010/2/17-2010/4/17	\\
4FGL J1549.5+0236	&	PKS 1546+027	&	237.3851	&	2.6084	&	2010/2/13-2010/2/20	&	2010/1/18-2010/3/18	\\
4FGL J1635.2+3808	&	4C +38.41	    &	248.8168	&	38.1401	&	2010/3/7	&	2010/2/7-2010/4/7	\\
4FGL J1640.4+3945	&	NRAO 512	    &	250.119	&	39.7626	&	2010/8/7	&	2010/7/7-2010/9/7	\\
4FGL J1709.7+4318	&	B3 1708+433	    &	257.4316	&	43.3109	&	2009/12/1	&	2009/11/1-2010/1/1	\\
4FGL J1733.0-1305	&	PKS 1730-13	    &	263.2632	&	-13.0858	&	2010/3/14	&	2010/2/26-2010/4/26	\\
4FGL J1848.4+3217	&	B2 1846+32A	    &	282.105	&	32.295	&	2010/10/6-2010/10/19	&	2010/9/7-2010/11/7	\\
4FGL J1911.2-2006	&	PKS B1908-201	&	287.8078	&	-20.1137	&	2009/10/4	&	2009/9/4-2009/11/4	\\
4FGL J1924.8-2914	&	PKS B1921-293	&	291.2136	&	-29.2468	&	2010/9/30	&	2010/8/30-2010/10/30	\\
4FGL J1958.0-3845	&	PKS 1954-388	&	299.5026	&	-38.7547	&	2010/4/9-2010/4/14	&	2010/3/12-2010/5/12	\\
4FGL J1959.1-4247	&	PMN J1959-4246	&	299.7963	&	-42.7852	&	2010/4/5-2010/4/14	&	2010/3/10-2010/5/10	\\
4FGL J2135.3-5006	&	PMN J2135-5006	&	323.8362	&	-50.1015	&	2010/4/22-2010/5/5	&	2010/3/29-2010/5/29	\\
4FGL J2145.0-3356	&	PMN J2145-3357	&	326.2533	&	-33.9439	&	2009/9/22-2009/9/24	&	2009/8/23-2009/10/23	\\
4FGL J2151.8-3027	&	PKS 2149-306	&	327.9655	&	-30.46	&	2010/5-4-2010/5/13	&	2010/4/8-2010/6/8	\\
4FGL J2157.5+3127	&	B2 2155+31	    &	329.3862	&	31.4552	&	2009/7/8-2009/7/12	&	2009/7/10-2009/9/10	\\
4FGL J2201.5-8339	&	PKS 2155-83	    &	330.3787	&	-83.6631	&	2010/7/5-2010/7/17	&	2010/6/10-2010/8/10	\\
4FGL J2212.0+2356	&	PKS 2209+236	&	333.0191	&	23.9334	&	2009/4/15-2009/4/21	&	2009/3/19-2009/5/19	\\
4FGL J2225.7-0457	&	3C 446	        &	336.4321	&	-4.9537	&	2010/5/22-2010/5/27	&	2010/4/24-2010/6/24	\\
4FGL J2253.9+1609	&	3C 454.3	    &	343.4963	&	16.1506	&	2009/12/4-2009/12/6	&	2009/11/5-2010/1/5	\\
4FGL J2258.1-2759	&	PKS 2255-282	&	344.5288	&	-27.9843	&	2010/5/20-2010/5/26	&	2010/4/22-2010/6/22	\\
4FGL J2321.9+3204	&	B2 2319+31	    &	350.4779	&	32.0737	&	2009/5/20	&	2009/4/20-2009/6/20	\\
4FGL J2327.5+0939	&	PKS 2325+093	&	351.8959	&	9.6543	&	2010/6/18-2010/6/21	&	2010/5/24-2010/7/24	\\
4FGL J2334.2+0736	&	TXS 2331+073	&	353.5573	&	7.602	&	2009/12/20	&	2009/11/20-2010/1/20	\\

\hline
\label{table1}
\end{tabular}
\\

\footnotetext[1]{The simultaneous date from infrared to X-ray band.}\footnotetext[2]{The integration time of the Fermi data. }
\end{table}
%==================================
\newpage
\begin{table}
\scriptsize
\caption{The Parameters Used to fit the SEDs\label{alltatatable}}
\tabcolsep 1.4mm
\begin{tabular}{ccccccccccccc}
\hline\hline

Fermi Name & Source Name & $z$  &$\rm logL_{BLR}~(\rm erg~s^{-1})$  & $\rm R~(10^{17}cm)$ & $\rm B~(\rm G)$ & $\delta$ & $s$ & $r$ & $N$  & $\gamma_{0}~(10^2)$ & $x~(\rm pc)$ & $\chi^{2}$ \\
~(1) & (2) & (3) & (4) & (5) & (6) & (7) & (8) & (9) & (10) & (11) & (12) & (13) \\
\hline
4FGL J0050.0-5736	&	PKS 0047-579	&	1.797	&	$44.87^a$			&	1.44	&	0.24	&	18	&	2.38	&	0.58	&	0.045	&	2.4	&	2.25	&	6.2	\\
4FGL J0051.1-0648	&	PKS 0048-071	&	1.975	&	$44.87^a$			&	4	&	0.24	&	15	&	2.33	&	0.58	&	0.012	&	2	&	0.38	&	69.11	\\
4FGL J0109.7+6133	&	TXS 0106+612	&	0.785	&	$44.87^a$			    &	1	&	0.19	&	17	&	2.24	&	0.5	&	0.006	&	1.7	&	1.9	&	24.71	\\
4FGL J0113.4+4948	&	S4 0110+49	    &	0.389	&	$44.87^a$			&	1.2	&	0.18	&	9.5	&	1.79	&	0.6	&	0.049	&	2	&	1.15	&	1.59	\\
4FGL J0205.0-1700	&	PKS 0202-17	    &	1.74	&	$44.87^a$			&	2.35	&	0.24	&	9	&	1.95	&	0.61	&	0.121	&	1.6	&	0.38	&	8.96	\\
4FGL J0217.4+7352	&	S5 0212+73	    &	2.367	&	$44.87^a$			    &	3.3	&	0.158	&	20	&	2.4	&	0.64	&	0.049	&	1.7	&	0.8	&	44.03	\\
4FGL J0217.8+0144	&	PKS 0215+015	&	1.715	&	$44.87^a$			&	1.65	&	0.329	&	24	&	2.24	&	0.76	&	0.025	&	2.1	&	1.95	&	22.08	\\
4FGL J0221.1+3556	&	S4 0218+35	    &	0.944	&	$44.87^a$			    &	1.9	&	0.147	&	15	&	1.85	&	0.77	&	0.013	&	3	&	4	&	31.41	\\
4FGL J0222.0-1616	&	PKS 0219-164	&	0.698	&	$44.87^a$			&	1.6	&	0.18	&	10	&	2	&	0.65	&	0.041	&	2	&	0.284	&	7.52	\\
4FGL J0236.8-6136	&	PKS 0235-618	&	0.465	&	$44.87^a$			&	0.55	&	0.49	&	11.1	&	1.98	&	0.83	&	0.18	&	2.3	&	0.384	&	38.29	\\
4FGL J0407.5+0741	&	TXS 0404+075	&	1.133	&	$44.51^b$			&	1.7	&	0.209	&	15	&	2.05	&	0.9	&	0.032	&	2	&	0.34	&	15.19	\\
4FGL J0438.4-1254	&	PKS 0436-129	&	1.276	&	$44.78^b$			&	2.65	&	0.22	&	12.5	&	2.05	&	0.58	&	0.029	&	1.4	&	0.27	&	10.81	\\
4FGL J0457.0-2324	&	PKS 0454-234	&	1.003	&	$44.87^a$			    &	2.81	&	0.364	&	12.11	&	2.18	&	0.62	&	0.009	&	2.2	&	0.8	&	12.11	\\
4FGL J0539.9-2839	&	PKS 0537-286	&	3.104	&	$44.87^a$			    &	4.95	&	0.115	&	23	&	2.28	&	0.87	&	0.017	&	1.5	&	1.5	&	24.95	\\
4FGL J0601.1-7035	&	PKS 0601-70	    &	2.409	&	$44.69^b$			    &	2.3	&	0.242	&	22.49	&	2.1	&	0.53	&	0.01	&	2	&	2.3	&	15.68	\\
4FGL J0654.4+4514	&	B3 0650+453	    &	0.933	&	$44.26^b$			&	1.5	&	0.227	&	12	&	1.87	&	0.62	&	0.058	&	1.8	&	0.68	&	11.74	\\
4FGL J0656.3-0322	&	TXS 0653-033	&	0.634	&	$45.68^b$			    &	2.354	&	0.17	&	12.08	&	2.35	&	0.54	&	0.008	&	2.3	&	3.9	&	22.91	\\
4FGL J0750.8+1229	&	OI 280	        &	0.889	&	$44.95^b$			&	2.37	&	0.26	&	12.82	&	2.25	&	0.65	&	0.075	&	1.52	&	1.85	&	97.5	\\
4FGL J0805.4+6147	&	TXS 0800+618	&	3.033	&	$44.56^b$			&	2.86	&	0.141	&	26.84	&	2	&	1.169	&	0.021	&	1.8	&	1.95	&	13.43	\\
4FGL J0920.9+4441	&	S4 0917+44	    &	2.19	&	$45.775^b$			    &	2.6	&	0.278	&	20	&	2.2	&	0.68	&	0.03	&	2	&	5.5	&	51.93	\\
4FGL J0956.7+2516	&	OK 290	        &	0.708	&	$44.93^b$			&	1.48	&	0.245	&	10	&	1.78	&	0.75	&	0.062	&	2	&	0.53	&	19.08	\\
4FGL J1037.7-2822	&	PKS B1035-281	&	1.066	&	$44.95^b$			    &	1.41	&	0.23	&	14.34	&	1.89	&	0.58	&	0.06	&	1.32	&	1.2	&	4.5	\\
4FGL J1106.0+2813	&	MG2 J110606+2812	&	0.844	&	$45.16^b$			&	2.03	&	0.22	&	10.9	&	1.78	&	0.835	&	0.008	&	3.2	&	0.69	&	7.22	\\
4FGL J1127.0-1857	&	 PKS 1124-186	&	1.048	&	$44.87^a$			    &	2.9	&	0.238	&	10.7	&	2.001	&	0.631	&	0.01	&	3.2	&	2.7	&	73.83	\\
4FGL J1147.0-3812	&	PKS 1144-379	&	1.048	&	$44.48^b$			&	2.4	&	0.303	&	13.78	&	2.35	&	0.66	&	0.021	&	2	&	0.25	&	12.53	\\
4FGL J1205.7-2635	&	PKS 1203-26	    &	0.789	&	$44.07^b$			&	1.84	&	0.143	&	15	&	2.02	&	0.59	&	0.005	&	2.6	&	0.45	&	6.42	\\
4FGL J1208.9+5441	&	TXS 1206+549	&	1.345	&	$44.52^b$			&	1.6	&	0.31	&	10	&	1.78	&	0.72	&	0.048	&	2	&	0.28	&	7.19	\\
4FGL J1222.5+0414	&	4C +04.42	    &	0.966	&	$44.97^c$			&	2.1	&	0.21	&	11.5	&	1.73	&	0.853	&	0.063	&	1.7	&	0.57	&	5.22	\\
4FGL J1246.7-2548	&	PKS 1244-255	&	0.635	&	$44.87^a$			    &	1.07	&	0.188	&	18.6	&	2.05	&	0.881	&	0.041	&	2.3	&	1.2	&	21.99	\\
4FGL J1256.1-0547	&	3C 279	        &	0.536	&	$44.78^b$			    &	3.05	&	0.34	&	13	&	2.35	&	0.68	&	0.033	&	1.5	&	1.6	&	353.3	\\
4FGL J1333.7+5056	&	CLASS J1333+5057	&	1.362	&	$44.37^b$			&	1	&	0.29	&	15.32	&	2.1	&	0.74	&	0.011	&	3	&	0.27	&	2.44	\\
4FGL J1347.6-3751	&	PMN J1347-3750	&	1.3	&	$44.67^b$			    &	0.8	&	0.371	&	13.6	&	2.07	&	0.646	&	0.085	&	2.2	&	0.38	&	2.2	\\
4FGL J1358.1+7642	&	S5 1357+76	    &	1.585	&	$44.2^b$			    &	1.2	&	0.18	&	20	&	2.08	&	0.63	&	0.029	&	2.3	&	0.8	&	2.5	\\
4FGL J1408.9-0751	&	PKS B1406-076	&	1.494	&	$45.47^b$			    &	1.55	&	0.293	&	17.6	&	1.85	&	0.98	&	0.034	&	1.93	&	3	&	46.22	\\
4FGL J1436.9+2321	&	PKS B1434+235	&	1.544	&	$44.72^b$			    &	2.26	&	0.16	&	19	&	1.9	&	1.24	&	0.004	&	3.3	&	0.6	&	11.98	\\
4FGL J1504.4+1029	&	PKS 1502+106	&	1.839	&	$45.24^b$			    &	2.2	&	0.204	&	16	&	2	&	0.96	&	0.014	&	3.32	&	4.3	&	46.94	\\
4FGL J1514.8+4448	&	NVSS J151436+445003	&	0.57	&	$43.33^b$			&	0.9	&	0.22	&	18	&	1.57	&	1.31	&	0.015	&	2	&	0.29	&	8.77	\\
4FGL J1539.6+2743	&	MG2 J153938+2744	&	2.19	&	$44.63^b$			&	1.36	&	0.197	&	20	&	2.1501	&	0.749	&	0.019	&	2.75	&	0.7	&	3.9	\\
4FGL J1549.5+0236	&	PKS 1546+027	&	0.414	&	$44.8^b$			&	1.2	&	0.325	&	8.5	&	2.1	&	0.59	&	0.08	&	1.88	&	0.238	&	34.56	\\
4FGL J1635.2+3808	&	4C +38.41	    &	1.814	&	$45.48^b$			    &	2.6	&	0.221	&	22	&	2.23	&	0.8	&	0.029	&	2	&	3.8	&	45.97	\\
4FGL J1640.4+3945	&	NRAO 512	    &	1.66	&	$45.01^c$			&	2	&	0.274	&	16	&	2.2	&	0.67	&	0.013	&	2.6	&	0.45	&	7.46	\\
4FGL J1709.7+4318	&	B3 1708+433	    &	1.027	&	$44.03^b$			    &	1.28	&	0.3	&	14	&	2	&	0.66	&	0.016	&	2.6	&	0.2	&	3.2	\\
4FGL J1733.0-1305	&	PKS 1730-13	    &	0.902	&	$44.83^b$			    &	2	&	0.278	&	20	&	2.2	&	0.7	&	0.019	&	1.7	&	0.5	&	89.23	\\
4FGL J1848.4+3217	&	B2 1846+32A	    &	0.798	&	$44.58^b$			&	0.79	&	0.3	&	15	&	2.31	&	0.702	&	0.055	&	3	&	0.62	&	59.06	\\
4FGL J1911.2-2006	&	PKS B1908-201	&	1.119	&	$44.87^a$			&	2.55	&	0.32	&	15.2	&	2.08	&	0.75	&	0.016	&	2	&	1.55	&	18	\\
4FGL J1924.8-2914	&	PKS B1921-293	&	0.352	&	$44.02^b$			&	3	&	0.27	&	10	&	2	&	0.93	&	0.021	&	2	&	0.85	&	42.37	\\
4FGL J1958.0-3845	&	PKS 1954-388	&	0.63	&	$44.2^b$			&	1.51	&	0.235	&	11	&	1.91	&	0.7	&	0.057	&	2	&	0.34	&	34.11	\\
4FGL J1959.1-4247	&	PMN J1959-4246	&	2.174	&	$45.13^b$			&	0.94	&	0.34	&	20	&	2	&	1	&	0.034	&	3	&	1.5	&	45.28	\\
4FGL J2135.3-5006	&	PMN J2135-5006	&	2.181	&	$45.26^b$			    &	1.2	&	0.21	&	21	&	2.1	&	1.02	&	0.053	&	2	&	0.9	&	2.97	\\
4FGL J2145.0-3356	&	PMN J2145-3357	&	1.36	&	$44.18^b$			    &	1.38	&	0.248	&	18.2	&	1.9	&	0.67	&	0.01	&	2.2	&	0.6	&	13.01	\\
4FGL J2151.8-3027	&	PKS 2149-306	&	2.345	&	$44.87^a$			&	1.6	&	0.62	&	7	&	2.01	&	0.85	&	0.68	&	1.75	&	0.15	&	13.21	\\
4FGL J2157.5+3127	&	B2 2155+31	    &	1.488	&	$44.74^b$			    &	1.14	&	0.28	&	18.75	&	2.18	&	0.72	&	0.048	&	2	&	1	&	7.36	\\
4FGL J2201.5-8339	&	PKS 2155-83	    &	1.865	&	$45.19^b$			    &	1.3	&	0.3	&	17	&	2.07	&	0.7	&	0.048	&	2	&	1.5	&	12.03	\\
4FGL J2212.0+2356	&	PKS 2209+236	&	1.125	&	$44.79^b$			&	1.2	&	0.28	&	15	&	1.8	&	1	&	0.035	&	2.2	&	0.65	&	6.33	\\
4FGL J2225.7-0457	&	3C 446	        &	1.404	&	$45.6^b$			&	3.62	&	0.335	&	13	&	2.22	&	0.7	&	0.015	&	2	&	0.28	&	33.79	\\
4FGL J2253.9+1609	&	3C 454.3	    &	0.859	&	$45.39^b$			&	3.1	&	0.28	&	25.5	&	2	&	1.285	&	0.0046	&	4	&	15.5	&	487.34	\\
4FGL J2258.1-2759	&	PKS 2255-282	&	0.926	&	$45.84^b$			    &	2.82	&	0.204	&	15	&	2.02	&	0.68	&	0.015	&	2	&	9.8	&	20.03	\\
4FGL J2321.9+3204	&	B2 2319+31	    &	1.489	&	$44.71^b$			&	1.85	&	0.204	&	17	&	1.85	&	0.73	&	0.015	&	2	&	1.65	&	31.59	\\
4FGL J2327.5+0939	&	PKS 2325+093	&	1.843	&	$45.2^b$			&	3.01	&	0.219	&	10	&	1.6	&	1.159	&	0.077	&	1.8	&	1.02	&	14.48	\\
4FGL J2334.2+0736	&	TXS 2331+073	&	0.401	&	$44.93^b$			    &	1.05	&	0.29	&	7.5	&	1.9	&	0.72	&	0.095	&	2.02	&	1	&	74.81	\\

\hline
\label{table2}
\end{tabular}
\\

\footnotetext[1]{The average broad-line luminosity.}\footnotetext[2]{The broad-line luminosity selected from \citet{2016MNRAS.463.3038X}.}\footnotetext[3]{The broad-line luminosity selected from \citet{2009MNRAS.397..985G}.}
\end{table}

%==================================
\newpage
\begin{table}
\scriptsize
\caption{The kinetic power in relativistic electrons, the power carried in magnetic field, the kinetic luminosity in cold protons, the radiative power, and the total jet power for FSRQs of our sample \label{alltatatable}}
\tabcolsep 1.4mm
\begin{tabular}{ccccccccccccc}
\hline\hline

Fermi Name & Source Name & $P_{\rm e}~(\rm erg\ s^{-1})$  &$P_{\rm B}~(\rm erg\ s^{-1})$  & $P_{\rm p}~(\rm erg\ s^{-1})$ & $P_{\rm r}~(\rm erg\ s^{-1})$ & $P_{\rm jet}~(\rm erg\ s^{-1})$  \\
~(1) & (2) & (3) & (4) & (5) & (6) & (7) \\
\hline
4FGL J0050.0-5736	&	PKS 0047-579	&	5.47E+45	&	1.45E+45	&	2.34E+47	&	3.30E+45	&	2.44E+47	\\
4FGL J0051.1-0648	&	PKS 0048-071	&	4.25E+45	&	7.77E+45	&	7.47E+46	&	4.29E+46	&	1.30E+47	\\
4FGL J0109.7+6133	&	TXS 0106+612	&	1.41E+45	&	3.52E+45	&	7.83E+46	&	1.84E+45	&	8.51E+46	\\
4FGL J0113.4+4948	&	S4 0110+49	&	7.11E+44	&	1.58E+44	&	1.14E+46	&	5.76E+44	&	1.28E+46	\\
4FGL J0205.0-1700	&	PKS 0202-17	&	3.69E+45	&	9.66E+44	&	9.81E+46	&	1.24E+46	&	1.15E+47	\\
4FGL J0217.4+7352	&	S5 0212+73	&	1.82E+46	&	4.08E+45	&	8.26E+47	&	4.78E+46	&	8.96E+47	\\
4FGL J0217.8+0144	&	PKS 0215+015	&	4.30E+45	&	6.36E+45	&	1.14E+47	&	6.53E+45	&	1.31E+47	\\
4FGL J0221.1+3556	&	S4 0218+35	&	2.28E+45	&	6.58E+44	&	2.36E+46	&	9.65E+44	&	2.75E+46	\\
4FGL J0222.0-1616	&	PKS 0219-164	&	1.08E+45	&	3.11E+44	&	2.36E+46	&	5.63E+45	&	3.07E+46	\\
4FGL J0236.8-6136	&	PKS 0235-618	&	8.08E+44	&	3.35E+44	&	1.25E+46	&	1.23E+45	&	1.49E+46	\\
4FGL J0407.5+0741	&	TXS 0404+075	&	1.82E+45	&	1.06E+45	&	3.38E+46	&	4.56E+45	&	4.12E+46	\\
4FGL J0438.4-1254	&	PKS 0436-129	&	1.65E+45	&	1.99E+45	&	4.53E+46	&	1.41E+46	&	6.30E+46	\\
4FGL J0457.0-2324	&	PKS 0454-234	&	1.34E+45	&	5.75E+45	&	3.11E+46	&	3.26E+45	&	4.14E+46	\\
4FGL J0539.9-2839	&	PKS 0537-286	&	1.16E+46	&	6.42E+45	&	3.98E+47	&	3.61E+46	&	4.52E+47	\\
4FGL J0601.1-7035	&	PKS 0601-70	&	3.07E+45	&	5.87E+45	&	9.27E+46	&	3.92E+45	&	1.06E+47	\\
4FGL J0654.4+4514	&	B3 0650+453	&	1.62E+45	&	6.26E+44	&	3.06E+46	&	1.77E+45	&	3.46E+46	\\
4FGL J0656.3-0322	&	TXS 0653-033	&	1.09E+45	&	8.76E+44	&	4.53E+46	&	8.01E+44	&	4.80E+46	\\
4FGL J0750.8+1229	&	OI 280	&	4.16E+45	&	2.34E+45	&	1.29E+47	&	3.04E+45	&	1.38E+47	\\
4FGL J0805.4+6147	&	TXS 0800+618	&	7.68E+45	&	4.39E+45	&	1.25E+47	&	3.95E+46	&	1.76E+47	\\
4FGL J0920.9+4441	&	S4 0917+44	&	8.27E+45	&	7.83E+45	&	1.95E+47	&	1.55E+46	&	2.27E+47	\\
4FGL J0956.7+2516	&	OK 290	&	1.35E+45	&	4.93E+44	&	1.90E+46	&	2.69E+45	&	2.36E+46	\\
4FGL J1037.7-2822	&	PKS B1035-281	&	1.16E+45	&	8.10E+44	&	2.73E+46	&	1.29E+45	&	3.06E+46	\\
4FGL J1106.0+2813	&	MG2 J110606+2812	&	9.41E+44	&	8.88E+44	&	7.87E+45	&	3.68E+45	&	1.34E+46	\\
4FGL J1127.0-1857	&	 PKS 1124-186	&	2.57E+45	&	2.04E+45	&	3.43E+46	&	4.19E+45	&	4.31E+46	\\
4FGL J1147.0-3812	&	PKS 1144-379	&	2.61E+45	&	3.76E+45	&	1.01E+47	&	1.07E+46	&	1.18E+47	\\
4FGL J1205.7-2635	&	PKS 1203-26	&	6.95E+44	&	5.84E+44	&	1.33E+46	&	1.35E+45	&	1.59E+46	\\
4FGL J1208.9+5441	&	TXS 1206+549	&	1.25E+45	&	9.22E+44	&	1.79E+46	&	3.98E+45	&	2.41E+46	\\
4FGL J1222.5+0414	&	4C +04.42	&	2.50E+45	&	9.64E+44	&	3.69E+46	&	3.71E+45	&	4.41E+46	\\
4FGL J1246.7-2548	&	PKS 1244-255	&	1.90E+45	&	5.25E+44	&	3.11E+46	&	2.13E+45	&	3.57E+46	\\
4FGL J1256.1-0547	&	3C 279	&	3.21E+45	&	6.81E+45	&	1.35E+47	&	2.75E+45	&	1.48E+47	\\
4FGL J1333.7+5056	&	CLASS J1333+5057	&	5.62E+44	&	7.40E+44	&	8.02E+45	&	1.40E+45	&	1.07E+46	\\
4FGL J1347.6-3751	&	PMN J1347-3750	&	1.26E+45	&	6.11E+44	&	2.86E+46	&	2.04E+45	&	3.25E+46	\\
4FGL J1358.1+7642	&	S5 1357+76	&	2.32E+45	&	6.99E+44	&	5.24E+46	&	2.87E+45	&	5.83E+46	\\
4FGL J1408.9-0751	&	PKS B1406-076	&	2.00E+45	&	2.39E+45	&	2.87E+46	&	2.94E+45	&	3.60E+46	\\
4FGL J1436.9+2321	&	PKS B1434+235	&	1.50E+45	&	1.77E+45	&	1.21E+46	&	7.10E+45	&	2.25E+46	\\
4FGL J1504.4+1029	&	PKS 1502+106	&	4.05E+45	&	1.93E+45	&	4.08E+46	&	4.08E+45	&	5.08E+46	\\
4FGL J1514.8+4448	&	NVSS J151436+445003	&	3.09E+44	&	4.76E+44	&	3.02E+45	&	2.38E+44	&	4.05E+45	\\
4FGL J1539.6+2743	&	MG2 J153938+2744	&	2.59E+45	&	1.08E+45	&	4.68E+46	&	5.47E+45	&	5.59E+46	\\
4FGL J1549.5+0236	&	PKS 1546+027	&	8.01E+44	&	4.12E+44	&	2.42E+46	&	2.88E+45	&	2.83E+46	\\
4FGL J1635.2+3808	&	4C +38.41	&	9.14E+45	&	5.99E+45	&	2.38E+47	&	1.67E+46	&	2.70E+47	\\
4FGL J1640.4+3945	&	NRAO 512	&	2.36E+45	&	2.88E+45	&	5.46E+46	&	1.12E+46	&	7.10E+46	\\
4FGL J1709.7+4318	&	B3 1708+433	&	8.87E+44	&	1.08E+45	&	1.48E+46	&	2.31E+45	&	1.91E+46	\\
4FGL J1733.0-1305	&	PKS 1730-13	&	2.25E+45	&	4.63E+45	&	7.27E+46	&	8.33E+45	&	8.79E+46	\\
4FGL J1848.4+3217	&	B2 1846+32A	&	1.86E+45	&	4.74E+44	&	4.18E+46	&	1.51E+45	&	4.57E+46	\\
4FGL J1911.2-2006	&	PKS B1908-201	&	2.31E+45	&	5.76E+45	&	5.17E+46	&	4.21E+45	&	6.39E+46	\\
4FGL J1924.8-2914	&	PKS B1921-293	&	1.47E+45	&	2.46E+45	&	2.62E+46	&	1.17E+45	&	3.13E+46	\\
4FGL J1958.0-3845	&	PKS 1954-388	&	1.57E+45	&	5.71E+44	&	2.82E+46	&	1.98E+45	&	3.24E+46	\\
4FGL J1959.1-4247	&	PMN J1959-4246	&	2.24E+45	&	1.53E+45	&	2.44E+46	&	3.95E+45	&	3.21E+46	\\
4FGL J2135.3-5006	&	PMN J2135-5006	&	2.86E+45	&	1.05E+45	&	6.13E+46	&	6.48E+45	&	7.17E+46	\\
4FGL J2145.0-3356	&	PMN J2145-3357	&	7.80E+44	&	1.45E+45	&	1.29E+46	&	1.53E+45	&	1.67E+46	\\
4FGL J2151.8-3027	&	PKS 2149-306	&	5.88E+45	&	1.81E+45	&	1.23E+47	&	9.18E+46	&	2.22E+47	\\
4FGL J2157.5+3127	&	B2 2155+31	&	2.20E+45	&	1.34E+45	&	5.97E+46	&	2.37E+45	&	6.56E+46	\\
4FGL J2201.5-8339	&	PKS 2155-83	&	2.33E+45	&	1.65E+45	&	5.46E+46	&	3.26E+45	&	6.18E+46	\\
4FGL J2212.0+2356	&	PKS 2209+236	&	1.16E+45	&	9.52E+44	&	1.37E+46	&	1.89E+45	&	1.77E+46	\\
4FGL J2225.7-0457	&	3C 446	&	3.43E+45	&	9.31E+45	&	1.02E+47	&	4.14E+47	&	5.29E+47	\\
4FGL J2253.9+1609	&	3C 454.3	&	8.41E+45	&	1.84E+46	&	6.04E+46	&	9.12E+45	&	9.63E+46	\\
4FGL J2258.1-2759	&	PKS 2255-282	&	2.70E+45	&	2.79E+45	&	5.96E+46	&	1.76E+45	&	6.69E+46	\\
4FGL J2321.9+3204	&	B2 2319+31	&	1.47E+45	&	1.54E+45	&	2.34E+46	&	2.05E+45	&	2.84E+46	\\
4FGL J2327.5+0939	&	PKS 2325+093	&	4.71E+45	&	1.63E+45	&	5.31E+46	&	9.55E+45	&	6.90E+46	\\
4FGL J2334.2+0736	&	TXS 2331+073	&	5.92E+44	&	1.95E+44	&	1.02E+46	&	3.69E+44	&	1.14E+46	\\

\hline
\label{table3}
\end{tabular}
\\

\end{table}

\end{document}